\documentclass{article}
\usepackage[latin9]{inputenc}
\usepackage{amsmath}
\usepackage{amssymb}
\usepackage{esint}

\makeatletter
\usepackage{amssymb}
\usepackage{amsfonts}
\usepackage{amsmath}
\usepackage{graphicx}
\usepackage{appendix}%
\providecommand{\U}[1]{\protect\rule{.1in}{.1in}}
\providecommand{\U}[1]{\protect\rule{.1in}{.1in}}
\providecommand{\U}[1]{\protect\rule{.1in}{.1in}}
\providecommand{\U}[1]{\protect\rule{.1in}{.1in}}
\newtheorem{theorem}{Theorem}

\newtheorem{corollary}[theorem]{Corollary}

\newtheorem{definition}[theorem]{Definition}
\newtheorem{example}[theorem]{Example}

\newtheorem{lemma}[theorem]{Lemma}
\newtheorem{notation}[theorem]{Notation}

\newtheorem{proposition}[theorem]{Proposition}
\newtheorem{remark}[theorem]{Remark}

\makeatother

\begin{document}

\title{The Non-Metricity Formulation of General Relativity\thanks{Published in \emph{Adv. Appl. Clifford Algebras.}}}

\author{Igor Mol\thanks{igormol@ime.unicamp.br or igormol@gmail.com.}\\
Institute of Mathematics, Statistics and Scientific Computation\\
Unicamp, SP, Brazil}
\maketitle
\begin{abstract}
After recalling the differential geometry of non-metric connections
in the formalism of differential forms, we introduce the idea of a
Non-Metricity (NM) connection, whose connection $1$--forms coincides
with the non-metricity $1$--forms for a class of cobase fields. Then
we formulate a theory of gravitation (equivalent to General Relativity
(GR)) which admits a geometrical interpretation in a flat torsionless
space where the gravitational field is completely manifest in the
non-metricity of a NM connection. We define and then apply the non-metricity
gauge to a gravitational Lagrangian density discovered by Wallner
\cite{W} (proved in Appendix \ref{EHL} to be equivalent to Einstein-Hilbert).
The Einstein equations coupled to the matter currents $\left(\mathcal{J}_{\alpha}\right)$
thus becomes $\delta dg_{\alpha}=\mathcal{T}_{\alpha}+\mathcal{J}_{\alpha}$,
where $\left(\mathcal{T}_{\alpha}\right)$ is identified as the gravitational
energy-momentum currents, to which we shall find a relatively simple
and physically appealing form. It is also shown that in the gravitational
analogue of the Lorenz gauge, our field equations can be written as
a system of Proca equations, which may be of interest in the study
of propagation of gravitational-electromagnetic waves.

\newpage{}
\end{abstract}
\tableofcontents{}

\section{Introduction}

In this paper, a theory of gravitation equivalent to General Relativity
(GR) will be formulated, for which a geometrical interpretation where
the gravitational field is manifest in the non-metricity of a flat
torsionless connection is naturally attributed.

In order to do so, it will be necessary to recall some facts about
the differential geometry of non-metric connections in parallelizable
manifolds, which are presented in sections \ref{Geometry} and \ref{UDF}
in the formalism of differential forms\footnote{In section \ref{UDF}, we also derive an identity decomposing the
connection $1$--forms of an arbitrary connection in terms of its
non-metricity $1$--forms, its torsion $2$--forms and some Levi-Civita
connection $1$--forms, something which may be useful to the study
of gravitational theories with additional degrees of freedom \cite{H}.}. Also, we introduce the concept of Non-Metricity (NM) connections,
having the property that its connection $1$--forms coincides with
its non-metricity $1$--forms relatively to a class of cobase fields,
as described in section \ref{NMCS}.

Subsequently to the mathematical preliminaries of section \ref{NMC},
we formulate our gravitational theory in section \ref{GNM}. In section
\ref{DFE}, we start from a gravitational Lagrangian density $\mathcal{L}$
discovered by Wallner \cite{W}, which is given in terms of a cobase
field $\left(g_{\alpha}\right)\in%TCIMACRO{\tbigwedge^{1}}%
%BeginExpansion
{\textstyle \bigwedge^{1}}%EndExpansion
M$ representing the gravitational potentials by
\[
\mathcal{L}=\frac{1}{2}g_{\alpha}\wedge dg^{\beta}\wedge\star\left(g_{\beta}\wedge dg^{\alpha}\right)-\frac{1}{4}g_{\alpha}\wedge dg^{\alpha}\wedge\star\left(g_{\beta}\wedge dg^{\beta}\right)\text{.}
\]
In Appendix \ref{EHL}, the equivalence between the Wallner Lagrangian
(WL) density and the Einstein-Hilbert Lagrangian density is established.

Then, also in section \ref{DFE}, we introduce the \textit{non-metricity
gauge}, whose geometrical meaning will become clear from the discussion
of NM connections presented in section \ref{NMCS}. It will be shown
from the variational principle (Proposition \ref{FIELD EQ}) that
the gravitational field equations assumes the form
\[
\delta dg_{\alpha}=\mathcal{T}_{\alpha}+\mathcal{J}_{\alpha}\text{,}
\]
where $\left(\mathcal{J}_{\alpha}\right)\in%TCIMACRO{\tbigwedge^{1}}%
%BeginExpansion
{\textstyle \bigwedge^{1}}%EndExpansion
M$ are the matter energy-momentum currents and $\left(\mathcal{T}_{\alpha}\right)\in%TCIMACRO{\tbigwedge^{1}}%
%BeginExpansion
{\textstyle \bigwedge^{1}}%EndExpansion
M$ are identified with the gravitational energy-momentum currents. In
the non-metricity gauge, we find a relatively\footnote{If compared, for instance, with \cite{R}.}
simple expression for the gravitational energy-momentum currents,
namely,
\begin{align*}
\mathcal{T}_{\alpha} & =\frac{1}{2}\star\left(dg_{\beta}\wedge i_{\alpha}\star dg^{\beta}-i_{\alpha}dg_{\beta}\wedge\star dg^{\beta}\right)\\
 & +\frac{1}{2}\delta g_{\beta}\wedge\delta g^{\beta}\wedge g_{\alpha}+i_{\beta}d\delta g^{\beta}\wedge g_{\alpha}-i_{\alpha}d\delta g^{\beta}\wedge g_{\beta}\text{,}
\end{align*}
whose physical significance is discussed in the many Remarks and Examples
of section \ref{DFE}.

In particular, we prove that if the gravitational Lorenz gauge $\delta g_{\alpha}=0$
($0\leq\alpha\leq3$) is adopted, the gravitational equations becomes
the following system of coupled Proca equations with variable mass,
\[
\square g_{\alpha}+\frac{1}{2}\left\langle dg_{\beta}|dg^{\beta}\right\rangle g_{\alpha}=i_{dg_{\beta}}\left(g_{\alpha}\wedge dg^{\beta}\right)+\mathcal{J}_{\alpha}\text{.}
\]
This latter form of the field equations may be of interest in the
study of the propagation of gravitational-electromagnetic waves, as
briefly outlined in Examples \ref{WAVES 1} and \ref{WAVES 2}.

As another straightforward application of the field equations, we
also derive a force law for the matter currents coupled to the gravitational
field. By identifying the $1$--form
\[
\mathcal{W}_{\xi}=\frac{1}{2}\star\left(dg_{\beta}\wedge i_{\xi}\star dg^{\beta}-i_{\xi}dg_{\beta}\wedge\star dg^{\beta}\right)
\]
with the gravitational energy-flow along a Killing vector field $\xi\in\sec TM$,
we easily prove that
\[
\delta\mathcal{W}_{\xi}=\left\langle i_{\xi}dg_{\alpha}|\mathcal{T}^{\alpha}+\mathcal{J}^{\alpha}\right\rangle \text{,}
\]
whose analogy with the Lorentz force law of electrodynamics is outlined
in Remark \ref{LORENTZ}.

In section \ref{NM}, we finally show that the gravitational field
equations can be completely rewritten in terms of the components $\mathbf{Q}_{\alpha\beta\gamma}$
of the non-metricity $2$--forms (defined in section \ref{Geometry})
of a NM connection, together with the matter energy-momentum currents,
becoming
\begin{align*}
i_{\nu}\mathcal{J}_{\mu} & =\mathbf{Q}_{\mu\lbrack\alpha\nu];}^{~~~~~~\alpha}+\mathbf{Q}_{\alpha\nu~;\mu}^{~~\alpha}-\left[\mathbf{Q}_{\alpha\beta}^{~~\alpha;\beta}+\frac{1}{2}\left(\mathbf{Q}_{\alpha\beta\gamma}\mathbf{Q}^{\alpha\lbrack\beta\gamma]}+\mathbf{Q}_{\alpha\beta}^{~~\alpha}\mathbf{Q}_{\gamma}^{~\beta\gamma}\right)\right]\eta_{\mu\nu}\\
 & -\mathbf{Q}_{\alpha\lbrack\mu\beta]}\mathbf{Q}^{\alpha\lbrack\beta\gamma]}\eta_{\gamma\nu}-\mathbf{Q}_{\mu\alpha\beta}\mathbf{Q}_{\nu}^{~[\alpha\beta]}-\mathbf{Q}_{\mu\lbrack\alpha\nu]}\mathbf{Q}_{~\beta}^{\alpha~\beta}\text{.}
\end{align*}
The gravitational field may therefore be interpreted as the manifestation
of the non-metricity of a flat and torsionless connection living in
the spacetime manifold.

\section{Non-Metric Connections\label{NMC}}

In section \ref{Geometry}, we state the basic definitions of differential
geometry in the language of differential forms and, in particular,
we discuss the non-metricity $1$--forms. Then, in section \ref{UDF},
many identities useful in the study of the gravitational field equations
and non-metricity are derived. Lastly, in section \ref{NMCS}, we
introduce the Non-Metricity (NM) connections in a parallelizable manifold,
as it will be employed in section \ref{DFE} to define the non-metricity
gauge.

\subsection{Basic Definitions \label{Geometry}}

On what follows, let $M$ be a parallelizable manifold.

\begin{notation} Here and thereafter, $(f_{\alpha\beta...})\in F$
signify a sequence $(f_{\alpha\beta...})_{\alpha\beta...}$ of elements
of the family $F$. Greek indexes always belongs to $\{0,1,...,\dim\left(M\right)-1\}$.
\end{notation}

\begin{definition} Recall that $\left(E_{\alpha}\right)\in\sec TM$
is called a \textit{base field} if every $X\in\sec TM$ can be written
as a linear combination of $\left(E_{\alpha}\right)$. On the other
hand, $\left(E_{\alpha}\right)$ is called a \textit{frame field}
if $\left(E_{\alpha}\right)$ is a base field orthonormal according
to a given metric \emph{\cite{Hicks} \cite{O'Neill}}. Also, $\left(g_{\alpha}\right)\in\sec T^{\ast}M$
is called the \textit{cobase field} of $\left(E_{\alpha}\right)$
if $g_{\alpha}\left(E_{\beta}\right)=\delta_{\alpha\beta}$ for all
$0\leq\alpha,\beta\leq\dim\left(M\right)-1$, and $\left(g_{\alpha}\right)$
is a \textit{coframe field }if $\left(E_{\alpha}\right)$ is a frame
field. For brevity, a frame field and a cobase field will be referred
to just as a \textit{frame} and a \textit{cobase}, respectively. \end{definition}

\begin{notation} $%TCIMACRO{\tbigwedge^{p}}%
%BeginExpansion
{\textstyle \bigwedge^{p}}%EndExpansion
M\equiv\sec%TCIMACRO{\tbigwedge^{p}}%
%BeginExpansion
{\textstyle \bigwedge^{p}}%EndExpansion
T^{\ast}M$ is the space of the sections of the bundle differential $p$--forms.
\end{notation}

\begin{notation} Let $\mathbf{g}$ be a metric tensor. We shall write
$g\left(X,Y\right)=\left\langle X|Y\right\rangle $ for all $X,Y\in\sec TM$.
\end{notation}

Let $\left(E_{\alpha}\right)$ a base field on $M$ and $\left(g_{\alpha}\right)$
its cobase field. The metric induced by $\left(g_{\alpha}\right)$
is $\mathbf{g}=\eta_{\alpha\beta}g^{\alpha}\otimes g^{\beta}$, and
the pair $\left(M,\mathbf{g}\right)$ is called a pseudo-Riemmanian
manifold. Here, 
\[
\left(\eta_{\alpha\beta}\right)=diag\left(-1,...,-1,+1,...,+1\right)
\]
posses $p$ negative and $q$ positive eigenvalues. Recall that that
$\mathbf{g}$ is called Lorentzian when $p=1$ or $q=1$, and we shall
adopt the former convention.

\begin{notation} We denote by $D$ an arbitrary connection on $M$,
while $\nabla$ is reserved to the Levi-Civita connection of $\left(M,\mathbf{g}\right)$.
The triple $\left(M,\mathbf{g},\nabla\right)$ is referred to as a
pseudo-Riemannian space. \end{notation}

Let $T$ be the torsion and $R$ the Riemman curvature tensor of $D$,
and $(\omega_{~\beta}^{\alpha})\in%TCIMACRO{\tbigwedge^{1}}%
%BeginExpansion
{\textstyle \bigwedge^{1}}%EndExpansion
M$ the connection $1$--forms of $D$ relative to the cobase $\left(g_{\alpha}\right)$.
Recall that for all $X\in\sec TM$,
\[
D_{X}E_{\alpha}=\omega_{~\alpha}^{\beta}(X)E_{\beta}\text{.}
\]

Since we shall use constantly the Levi-Civita connection to apply
some tricks of differential forms, we must distinguish its connection
$1$--forms to that of an arbitrary connection. So we introduce the
following notation.

\begin{notation} The connection $1$--forms of the Levi-Civita connection
$\nabla$ will be denoted by $(\theta_{~\beta}^{\alpha})$, and they
will be called the Levi-Civita connection $1$--forms. \end{notation}

Also, let $\left(\mathbf{T}^{\alpha}\right)$ and $(\mathcal{R}_{~\beta}^{\alpha})$
$\in%TCIMACRO{\tbigwedge^{2}}%
%BeginExpansion
{\textstyle \bigwedge^{2}}%EndExpansion
M$ be the torsion and curvature $2$--forms relative to the cobase $\left(g_{\alpha}\right)$,
and recall that for all $X,Y\in\sec TM$,
\[
T(X,Y)=\mathbf{T}^{\alpha}\left(X,Y\right)E_{\alpha}\text{,}~~~R(X,Y)E_{\alpha}=\mathcal{R}_{~\alpha}^{\beta}\left(X,Y\right)E_{\beta}\text{.}
\]

The Ricci tensor $Ric\in\sec T_{2}^{0}M$ of $D$ is defined to be
(up to an arbitrary sign) the contraction of $R$ such that, for all
$X,Y\in\sec TM$, $Ric\left(X,Y\right)=g^{\alpha}\left[R\left(X,E_{\alpha}\right)Y\right]$.
The Ricci $1$--forms $\left(\mathcal{R}_{\alpha}\right)$ are given
by $X\in\sec TM\mapsto\mathcal{R}_{\alpha}\left(X\right)=Ric\left(E_{\alpha},X\right)$.
Therefore, 
\begin{align*}
i_{\beta}\mathcal{R}_{\alpha} & =g^{\gamma}\left[R\left(E_{\beta},E_{\gamma}\right)E_{\alpha}\right]=g^{\gamma}\left[\mathcal{R}_{~\alpha}^{\delta}\left(E_{\beta},E_{\gamma}\right)E_{\delta}\right]\\
 & =\mathcal{R}_{~\alpha}^{\gamma}\left(E_{\beta},E_{\gamma}\right)=-i_{\beta}i_{\gamma}\mathcal{R}_{~\alpha}^{\gamma}\text{.}
\end{align*}
Since this holds for all $\beta$, we have just proven that
\begin{equation}
\mathcal{R}_{\alpha}=-i_{\beta}\mathcal{R}_{~\alpha}^{\beta}\text{.}\label{Ricci}
\end{equation}

One can prove using the above definitions that the Cartan structural
equations holds for an arbitrary connection $D$ \cite{Hicks}. We
shall, only for the sake of organization, state these equations as
a Lemma.

\begin{lemma} The connection $1$--forms $(\omega_{~\beta}^{\alpha})$
and the torsion and curvature $2$--forms $\left(\mathbf{T}^{\alpha}\right)$
and $(\mathcal{R}_{~\beta}^{\alpha})$ of $D$ relative to $\left(g_{\alpha}\right)$
obeys the Cartan structural equations
\begin{equation}
dg_{\alpha}=-\omega_{a}^{~\beta}\wedge g_{\beta}+\mathbf{T}_{\alpha}\text{,}
\end{equation}
\begin{equation}
\mathcal{R}_{~\beta}^{\alpha}=d\omega_{~\beta}^{\alpha}+\omega_{~\gamma}^{\alpha}\wedge\omega_{~\beta}^{\gamma}\text{.}
\end{equation}

\end{lemma}

Given the recurrent use of the first Cartan structural equation in
the following developments, for the sake of brevity, it is worth to
introduce the next notation.

\begin{notation} $\mathcal{G}_{\alpha}=dg_{\alpha}-\mathbf{T}_{a}\in%TCIMACRO{\tbigwedge^{2}}%
%BeginExpansion
{\textstyle \bigwedge^{2}}%EndExpansion
M$, $0\leq\alpha\leq\dim\left(M\right)-1$. \end{notation}

So now the first Cartan structural equation becomes just $\mathcal{G}_{\alpha}=-\omega_{a}^{~\beta}\wedge g_{\beta}$.

The following definition of the non-metricity $1$--forms $\left(\mathcal{A}_{\alpha\beta}\right)$
of $D$ relative to the cobase $\left(g_{\alpha}\right)$ is nonstandard
(besides its inherent simplicity), so we shall be more careful in
its formulation.

\begin{definition} \label{Def. NM}\label{Non-Metricity Field}The
non-metricity $1$--forms $(\mathcal{A}_{\alpha\beta})\in%TCIMACRO{\tbigwedge^{1}}%
%BeginExpansion
{\textstyle \bigwedge^{1}}%EndExpansion
M$ of $D$ relative to the cobase $\left(g_{\alpha}\right)$ are such
that
\[
\mathcal{A}_{\alpha\beta}\left(X\right)=-\frac{1}{2}\left(D_{X}g\right)\left(E_{\alpha},E_{\beta}\right)
\]
for all $X\in\sec TM$, while the non-metricity $2$--forms $\left(\mathbf{Q}_{\gamma}\right)\in%TCIMACRO{\tbigwedge^{2}}%
%BeginExpansion
{\textstyle \bigwedge^{2}}%EndExpansion
M$ are given by
\[
\mathbf{Q}_{\gamma}=\frac{1}{2}\mathbf{Q}_{\alpha\beta\gamma}g^{\alpha}\wedge g^{\beta}\text{,}
\]
where
\[
\mathbf{Q}_{\alpha\beta\gamma}=i_{[\alpha}\mathcal{A}_{\beta]\gamma}\equiv i_{\alpha}\mathcal{A}_{\beta\gamma}-i_{\beta}\mathcal{A}_{\alpha\gamma}\text{.}
\]

\end{definition}

\begin{lemma} \label{Non-metricity}The non-metricity $1$--forms
$(\mathcal{A}_{~\beta}^{\alpha})$ are the symmetric part of the connection
$1$--forms $(\omega_{~\beta}^{\alpha})$, that is,
\[
\mathcal{A}_{\alpha\beta}\mathcal{=}\frac{\omega_{\alpha\beta}+\omega_{\beta\alpha}}{2}\text{.}
\]

\end{lemma}

In fact, $0=D_{X}\left[g\left(E_{\alpha},E_{\beta}\right)\right]=\left(D_{X}g\right)\left(E_{\alpha},E_{\beta}\right)+\left\langle D_{X}E_{\alpha}|E_{\beta}\right\rangle +\left\langle E_{\alpha}|D_{X}E_{\beta}\right\rangle =-2\mathcal{A}_{\alpha\beta}\left(X\right)+\left\langle \omega_{~\alpha}^{\gamma}(X)E_{\gamma}|E_{\beta}\right\rangle +\left\langle E_{\alpha}|\omega_{~\beta}^{\gamma}(X)E_{\gamma}\right\rangle =-2\mathcal{A}_{\alpha\beta}\left(X\right)+\omega_{\alpha\beta}+\omega_{\beta\alpha}$,
proving our Lemma. $\blacksquare$

\bigskip{}

We conclude this section by introducing a notation for the anti-symmetric
part of $(\omega_{~\beta}^{\alpha})$, which will be related in an
important identity to the connection $1$--forms $(\theta_{~\beta}^{\alpha})$
of the Levi-Civita connection.

\begin{definition} \label{anti-symm.}Let $(\omega_{~\beta}^{\alpha})$
be the connection $1$--forms of some connection in $M$. The anti-symmetric
part $\left(\omega_{\left[\alpha\beta\right]}\right)$ of $(\omega_{~\beta}^{\alpha})$
are the $1$--forms given by 
\[
\omega_{\left[\alpha\beta\right]}=\frac{1}{2}\left(\omega_{\alpha\beta}-\omega_{\beta\alpha}\right)
\]
for all\emph{ }$0\leq\alpha,\beta\leq\dim\left(M\right)-1$. \end{definition}

Observe that, in the above notation, $\omega_{\alpha\beta}=\omega_{\left[\alpha\beta\right]}+\mathcal{A}_{\alpha\beta}$.

\subsection{Useful Identities \label{UDF}}

Now we prove some identities\ which are useful to study the gravitational
equations in the formalism of differential forms. We also derive the
decomposition formula of the connection $1$--forms of an arbitrary
connection in $M$ in terms of its non-metricity $1$--forms, its
torsion $2$--forms and the Levi-Civita connection $1$-forms.

\bigskip{}

The following Lemma gives an identity involving the codifferential
of a cobase and the Levi-Civita connection\textit{ }$1$--forms relative
to that cobase. It will be used only as an intermediary step for the
proof of the succeeding Lemmas. As the Levi-Civita connection always
exists in the pseudo-Riemannian space $\left(M,\mathbf{g}\right)$,
its use will not imply in any no loss of generality.

\begin{lemma} \label{Codiff.}The codifferential $\delta g_{\alpha}$
and the Levi-Civita connection $1$--forms $(\theta_{~\beta}^{\alpha})$
relative to the cobase $\left(g_{\alpha}\right)$ are related by
\[
\delta g_{\alpha}=i^{\beta}\theta_{\beta\alpha}\text{.}
\]

\end{lemma}

To prove this, we shall need the following identity
\[
d\star g_{\alpha}=-\theta_{\alpha\beta}\wedge\star g^{\beta}\text{.}
\]
Indeed, let $\varepsilon_{\alpha\beta\gamma\delta}$ be the Levi-Civita
totally anti-symmetric symbol. So
\begin{align*}
d\star g_{\alpha} & =\frac{1}{3!}\varepsilon_{\alpha\beta\gamma\delta}d\left(g^{\beta}\wedge g^{\gamma}\wedge g^{\delta}\right)=dg^{\beta}\wedge\frac{1}{2}\varepsilon_{\alpha\beta\gamma\delta}g^{\gamma}\wedge g^{\delta}\\
 & =dg^{\beta}\wedge\star\left(g_{\alpha}\wedge g_{\beta}\right)=-\theta_{~\gamma}^{\beta}\wedge\star\left(g_{\alpha}\wedge g_{\beta}\right)\wedge g^{\gamma}\text{.}
\end{align*}
On the other hand,
\begin{align*}
\star\left(g_{\alpha}\wedge g_{\beta}\right)\wedge g^{\gamma} & =\star^{2}\left(\star\left(g_{\alpha}\wedge g_{\beta}\right)\wedge g^{\gamma}\right)=-\star i^{\gamma}\left(g_{\alpha}\wedge g_{\beta}\right)\\
 & =-\star\left(\delta_{\alpha}^{\gamma}g_{\beta}-\delta_{\beta}^{\gamma}g_{\alpha}\right)\text{.}
\end{align*}
Therefore $d\star g_{\alpha}=\theta_{~\gamma}^{\beta}\wedge\star\left(\delta_{\alpha}^{\gamma}g_{\beta}-\delta_{\beta}^{\gamma}g_{\alpha}\right)=\theta_{~\alpha}^{\beta}\wedge\star g_{\beta}-\theta_{~\beta}^{\beta}\wedge\star g_{\alpha}=-\theta_{\alpha\beta}\wedge\star g^{\beta}$.

Now we can establish the Lemma. Just recall that 
\begin{align*}
\delta g_{\alpha} & =-\star\left(d\star g_{\alpha}\right)=\star\left(\theta_{\alpha\beta}\wedge\star g^{\beta}\right)\\
 & =\star\left(\star g^{\beta}\wedge\theta_{\beta\alpha}\right)=i_{\theta_{\beta\alpha}}\star^{2}g^{\beta}=i^{\beta}\theta_{\beta\alpha}.\blacksquare
\end{align*}

\bigskip{}

The next Lemma utilizes the last one to give a connection-independent
result, i.e., which holds in any parallelizable manifold\footnote{Even if Lemma \ref{Codiff.} employs a Levi-Civita connection, recall
that in any parallelizable manifold we can construct a pseudo-Riemannian
space by choosing a frame field and declaring it to be orthonormal.}.

\begin{lemma} \label{Trick}The contraction of the differential of
a cobase is related to the codifferential of that cobase by
\[
i^{\alpha}dg_{\alpha}=-\delta g^{\alpha}\wedge g_{\alpha}\text{.}
\]
In particular, it follows that
\[
i^{\alpha}dg_{\alpha}\wedge\star i^{\beta}dg_{\beta}=\delta g^{\alpha}\wedge\star\delta g_{\alpha}\text{.}
\]

\end{lemma}

First, $i^{\alpha}dg_{\alpha}=i^{\alpha}\left(-\theta_{\alpha\beta}\wedge g^{\beta}\right)=-i^{\alpha}\theta_{\alpha\beta}\wedge g^{\beta}+\theta_{\alpha\beta}g^{\alpha\beta}=-\delta g_{\beta}\wedge g^{\beta}$.

Second, using that $g^{\alpha}\wedge\star g^{\beta}=\star^{2}\left(\star g^{\beta}\wedge g^{\alpha}\right)=\star i^{\alpha}g^{\beta}=g^{\alpha\beta}\star1$,
\begin{align*}
i^{\alpha}dg_{\alpha}\wedge\star i^{\beta}dg_{\beta} & =\delta g^{\alpha}\wedge g_{\alpha}\wedge\star\left(\delta g^{b}\wedge g_{b}\right)\\
 & =\delta g^{\alpha}\wedge\delta g^{b}\wedge g_{\alpha}\wedge\star g_{b}\\
 & =\delta g^{\alpha}\wedge\delta g^{b}\eta_{\alpha\beta}\star1\\
 & =\delta g^{\alpha}\wedge\star\delta g_{\alpha}\text{. }\blacksquare
\end{align*}

The identity of the following Lemma is the bridge between the geometry
of an arbitrary connection in $M$ to the geometry of the Levi-Civita
connection of $\left(M,\mathbf{g}\right)$.

\begin{lemma} \label{Decomp. formula}The anti-symmetric part $\omega_{\left[\alpha\beta\right]}$
of the connection $1$--forms $(\omega_{~\beta}^{\alpha})$ of $D$
relative to the cobase $\left(g_{\alpha}\right)$ are given in terms
of $\mathcal{G}_{\alpha}=dg_{\alpha}-\mathbf{T}_{\alpha}$ \emph{(recall
that }$\left(\mathbf{T}_{\alpha}\right)$\emph{ are the torsion }$2$\emph{--forms
of }$D$\emph{)} and of the components $\mathbf{Q}_{\alpha\beta\gamma}$
of the non-metricity $2$--forms $\mathbf{Q}_{\gamma}$ of $D$ by
\[
\omega_{\left[\alpha\beta\right]}=i_{\beta}\mathcal{G}_{\alpha}-i_{\alpha}\mathcal{G}_{b}+\frac{1}{2}i_{\alpha}i_{\beta}\left(g_{\gamma}\wedge\mathcal{G}^{\gamma}\right)-\mathbf{Q}_{\alpha\beta\gamma}\wedge g^{\gamma}\text{.}
\]

\end{lemma}

Indeed, recall the first Cartan structural equation, $\mathcal{G}_{\gamma}=dg_{\gamma}-\mathbf{T}_{\gamma}=-\omega_{\gamma\delta}\wedge g^{\delta}$.
Contraction gives $i_{\beta}\mathcal{G}_{\gamma}=-i_{\beta}\omega_{\gamma\delta}\wedge g^{\delta}+\omega_{\gamma\beta}$.
Repeated contraction together with cyclic permutations yields \begin{subequations}
\label{0}
\begin{align}
i_{\alpha}i_{\beta}\mathcal{G}_{\gamma} & =i_{\alpha}\omega_{\gamma\beta}-i_{\beta}\omega_{\gamma\alpha}\text{,}\label{Cycle 1}\\
i_{\gamma}i_{\alpha}\mathcal{G}_{\beta} & =i_{\gamma}\omega_{\beta\alpha}-i_{\alpha}\omega_{\beta\gamma}\text{,}\label{Cycle 2}\\
i_{\beta}i_{\gamma}\mathcal{G}_{\alpha} & =i_{\beta}\omega_{\alpha\gamma}-i_{\gamma}\omega_{\alpha\beta}\text{.}\label{Cycle 3}
\end{align}
After summing Eqs.(\ref{Cycle 2}) and (\ref{Cycle 3}), multiplying
by $-1$ and interchanging $\beta$ with $\gamma$ in $i_{\beta}i_{\gamma}\mathcal{G}_{\alpha}$,
we obtain \end{subequations} 
\[
i_{\gamma}\left(\omega_{\alpha\beta}-\omega_{\beta\alpha}\right)-i_{\beta}\omega_{\alpha\gamma}+i_{\alpha}\omega_{\beta\gamma}=i_{\gamma}i_{\beta}\mathcal{G}_{\alpha}-i_{\gamma}i_{\alpha}\mathcal{G}_{\beta}\text{.}
\]
Now, summing with Eq.(\ref{Cycle 1}), dividing by $2$ and using
the definitions of\ $\omega_{\left[\alpha\beta\right]}$ (see Notation
\ref{anti-symm.}) and $\mathcal{A}_{\alpha\beta}$ (see Lemma \ref{Non-metricity})
yields
\begin{equation}
i_{\gamma}\omega_{\left[\alpha\beta\right]}+i_{\alpha}\mathcal{A}_{\beta\gamma}-i_{\beta}\mathcal{A}_{\alpha\gamma}=\frac{1}{2}\left(i_{\gamma}i_{\beta}\mathcal{G}_{\alpha}-i_{\gamma}i_{\alpha}\mathcal{G}_{\beta}+i_{\alpha}i_{\beta}\mathcal{G}_{\gamma}\right)\text{.}\label{Partial Sum}
\end{equation}
To remove the contraction $i_{\gamma}$, observe that 
\begin{align*}
i_{\alpha}\mathcal{A}_{\beta\gamma}-i_{\beta}\mathcal{A}_{\alpha\gamma} & =\delta_{\gamma}^{\delta}\left(i_{\alpha}\mathcal{A}_{\beta\delta}-i_{\beta}\mathcal{A}_{\alpha\delta}\right)\\
 & =i_{\gamma}\left(g^{\delta}\wedge\left(i_{\alpha}\mathcal{A}_{\beta\delta}-i_{\beta}\mathcal{A}_{\alpha\delta}\right)\right)\\
 & =i_{\gamma}\left(g^{\delta}\wedge\mathbf{Q}_{\alpha\beta\delta}\right)\text{,}
\end{align*}
and that $i_{\alpha}i_{\beta}\mathcal{G}_{\gamma}=\delta_{\gamma}^{\delta}i_{\alpha}i_{\beta}\mathcal{G}_{\delta}=i_{\gamma}\left(g^{\delta}\wedge i_{\alpha}i_{\beta}\mathcal{G}_{\delta}\right)$.
Therefore, Eq.(\ref{Partial Sum}) becomes
\[
i_{\gamma}\omega_{\left[\alpha\beta\right]}=i_{\gamma}\left[\frac{1}{2}\left(i_{\beta}\mathcal{G}_{\alpha}-i_{\alpha}\mathcal{G}_{\beta}+g^{\delta}\wedge i_{\alpha}i_{\beta}\mathcal{G}_{\delta}\right)-\mathbf{Q}_{\alpha\beta\delta}\wedge g^{\delta}\right]\text{.}
\]
Lastly, since this holds for any contraction $i_{\gamma}$ and
\begin{align}
g^{\delta}\wedge i_{\alpha}i_{\beta}\mathcal{G}_{\delta} & =-i_{\alpha}\left(g^{\delta}\wedge i_{\beta}\mathcal{G}_{\delta}\right)+i_{\beta}\mathcal{G}_{\alpha}\nonumber \\
 & =-i_{\alpha}\left(-i_{\beta}\left(g^{\delta}\wedge\mathcal{G}_{\delta}\right)+\mathcal{G}_{\beta}\right)+i_{\beta}\mathcal{G}_{\alpha}\nonumber \\
 & =i_{\alpha}i_{\beta}\left(g^{\delta}\wedge\mathcal{G}_{\delta}\right)-i_{\alpha}\mathcal{G}_{\beta}+i_{\beta}\mathcal{G}_{\alpha}\text{,}\label{CONTR}
\end{align}
we obtain the final form:
\[
\omega_{\left[\alpha\beta\right]}=i_{\beta}\mathcal{G}_{\alpha}-i_{\alpha}\mathcal{G}_{\beta}+\frac{1}{2}i_{\alpha}i_{\beta}\left(g^{\delta}\wedge\mathcal{G}_{\delta}\right)-\mathbf{Q}_{\alpha\beta\delta}\wedge g^{\delta}\text{. }\blacksquare
\]

Because the non-metricity $1$--forms and the torsion $2$--forms
vanishes in the Levi-Civita connection, the last Lemma applied to
the Levi-Civita connection $1$--forms $(\theta_{~\beta}^{\alpha})$
yields
\begin{equation}
\theta_{\alpha\beta}=i_{\beta}dg_{\alpha}-i_{\alpha}dg_{\beta}+\frac{1}{2}i_{\alpha}i_{\beta}\left(g_{\gamma}\wedge dg^{\gamma}\right)\text{,}\label{Levi-Civita}
\end{equation}
(recall that $\theta_{\alpha\beta}=$ $\theta_{\left[\alpha\beta\right]}$
for $\nabla$). The above equation is the correspondent of the Christoffel
symbols of the classical tensor calculus, and express the fact that
the Levi-Civita connection is completely determined by the cobase
$\left(g_{\alpha}\right)$.

As a result, we can finally derive our decomposition formula.

\begin{corollary} \label{Decomposition}The connection $1$--forms
$(\omega_{~\beta}^{\alpha})$ of an arbitrary connection $D$ is given
in terms of the Levi-Civita connection $1$--forms $(\theta_{~\beta}^{\alpha})$,
the non-metricity\footnote{Recall that $\mathbf{Q}_{\alpha\beta\gamma}$ are just the components
of the non-metricity $2$--forms $\mathbf{Q}_{\gamma}$, so that the
term ``$\mathbf{Q}_{\alpha\beta\gamma}\wedge g^{\gamma}$\textquotedblright \ is
another contribution of the non-metricity $1$--forms to $\omega_{\alpha\beta}$.} $1$--forms $(\mathcal{A}_{~\beta}^{\alpha})$ and torsion $2$--forms
$(\mathbf{T}^{\alpha})$ of $D$ \emph{(all relative to a fixed cobase
}$\left(g_{\alpha}\right)$\emph{)} via
\[
\omega_{\alpha\beta}=\theta_{\alpha\beta}+\mathcal{A}_{\alpha\beta}-\mathfrak{T}_{\alpha\beta}-\mathbf{Q}_{\alpha\beta\gamma}\wedge g^{\gamma}\text{,}
\]
where$\mathfrak{\ T}_{\alpha\beta}=\mathfrak{T}_{\alpha\beta}\left(\mathbf{T}_{\gamma}\right)\in%TCIMACRO{\tbigwedge^{1}}%
%BeginExpansion
{\textstyle \bigwedge^{1}}%EndExpansion
M$ is given by
\[
\mathfrak{T}_{\alpha\beta}=i_{\beta}\mathbf{T}_{\alpha}-i_{\alpha}\mathbf{T}_{\beta}+\frac{1}{2}i_{\alpha}i_{\beta}\left(g_{\gamma}\wedge\mathbf{T}^{\gamma}\right)\text{,}
\]
for all $0\leq\alpha,\beta\leq\dim\left(M\right)-1$. \end{corollary}

In fact, from the decomposition formula of Lemma \ref{Decomp. formula}
and $\mathcal{G}_{\alpha}=dg_{\alpha}-\mathbf{T}_{\alpha}$,
\begin{align*}
\omega_{\left[\alpha\beta\right]} & =i_{\beta}dg_{\alpha}-i_{\alpha}dg_{\beta}+\frac{1}{2}i_{\alpha}i_{\beta}\left(g_{\gamma}\wedge dg^{\gamma}\right)\\
 & -\left[i_{\beta}\mathbf{T}_{\alpha}-i_{\alpha}\mathbf{T}_{\beta}+\frac{1}{2}i_{\alpha}i_{\beta}\left(g_{\gamma}\wedge\mathbf{T}^{\gamma}\right)\right]\\
 & -\mathbf{Q}_{\alpha\beta\delta}\wedge g^{\delta}\text{.}
\end{align*}
Recognizing that the first sum is just the Levi-Civita connection
$1$--form $\theta_{\alpha\beta}$ (see Eq.(\ref{Levi-Civita})) and
that the second sum is what we called $\mathfrak{T}_{\alpha\beta}$,
we conclude that $\omega_{\left[\alpha\beta\right]}=\theta_{\alpha\beta}+\mathfrak{T}_{\alpha\beta}-\mathbf{Q}_{\alpha\beta\delta}\wedge g^{\delta}$.
Now, just recall that $\omega_{\alpha\beta}=\omega_{\left[\alpha\beta\right]}+\mathcal{A}_{\alpha\beta}$,
the sum of its symmetric and anti-symmetric parts. $\blacksquare$

\begin{remark} In the decomposition formula of $\omega_{\alpha\beta}$,
the contribution of the torsion derives from the $1$--forms $(\mathfrak{T}_{\alpha\beta})$,
which have the same structure as \emph{Eq.(\ref{Levi-Civita})} for
the Levi-Civita connection $1$--forms $(\theta_{~\beta}^{\alpha})$.
This symmetry between $\theta_{\alpha\beta}$ and $\mathfrak{T}_{\alpha\beta}$\emph{
}is relevant to the teleparallel formulation of GR. In fact, let $\left(g_{\alpha}\right)$
be a teleparallel cobase\footnote{That is, $D_{E_{\alpha}}E_{\beta}=0$ for all $0\leq\alpha,\beta\leq3$,
where $\left(E_{\alpha}\right)$ is the dual base field of $\left(g_{\alpha}\right)$.} in a teleparallel space with connection $D$. Thus $\mathbf{T}_{\alpha}=dg_{\alpha}$,
in which case $\mathfrak{T}_{\alpha\beta}$ reduces to the right-hand
side of \emph{Eq.(\ref{Levi-Civita})}. Hence, the torsion $2$--forms
in the teleparallel space $(M,g,D)$ are \emph{(up to a gauge transformation}
$dg_{\alpha}\mapsto dg_{\alpha}+d\chi$\emph{)} in a one-to-one correspondence
with connection $1$--forms in the pseudo-Riemannian space $(M,g,\nabla)$.
Therefore, if a geometrical theory of gravity can be formulated in
one space, it can be in the other \cite{H} \cite{P} \cite{M}. \end{remark}

We finish this section with another application of Lemma \ref{Decomp. formula},
proving a formula useful in the decomposition of the Einstein $1$--forms
(Eq.(\ref{EINST1-FORM})).

\begin{lemma} \label{Trick_}$d\star\left(g^{\alpha}\wedge g^{\beta}\right)=\delta g^{\beta}\wedge\star g^{\alpha}-\delta g^{\alpha}\wedge\star g^{\beta}+g^{\beta}\wedge\star dg^{\alpha}-g^{\alpha}\wedge\star dg^{\beta}-\star i^{\alpha}i^{\beta}\left(g_{\gamma}\wedge dg^{\gamma}\right)$.
\end{lemma}

Indeed, let $(\theta_{~\beta}^{\alpha})$ be the Levi-Civita connection
$1$--forms of $\left(M,\mathbf{g}\right)$ with $\mathbf{g}$ induced
by $\left(g_{\alpha}\right)$. So,
\begin{align*}
d\star\left(g^{\alpha}\wedge g^{\beta}\right) & =-\theta_{~\gamma}^{\alpha}\wedge\star\left(g^{\gamma}\wedge g^{\beta}\right)-\theta_{~\gamma}^{\beta}\wedge\star\left(g^{\alpha}\wedge g^{\gamma}\right)\\
 & =\theta_{~\gamma}^{\alpha}\wedge i^{\gamma}\star g^{\beta}-\theta_{~\gamma}^{\beta}\wedge i^{\gamma}\star g^{\alpha}\text{.}
\end{align*}
Using well-know properties of contraction and Lemma \ref{Codiff.},
\begin{align}
d\star\left(g^{\alpha}\wedge g^{\beta}\right) & =-i^{\gamma}(\theta_{~\gamma}^{\alpha}\wedge\star g^{\beta})+i^{\gamma}\theta_{~\gamma}^{\alpha}\wedge\star g^{\beta}+i^{\gamma}(\theta_{~\gamma}^{\beta}\wedge\star g^{\alpha})-i^{\gamma}\theta_{~\gamma}^{\beta}\wedge\star g^{\alpha}\nonumber \\
 & =i^{\gamma}(\star\theta_{~\gamma}^{\alpha}\wedge g^{\beta})-i^{\gamma}(\star\theta_{~\gamma}^{\beta}\wedge g^{\alpha})-i^{\gamma}\theta_{\gamma}^{~\alpha}\wedge\star g^{\beta}+i^{\gamma}\theta_{\gamma}^{~\beta}\wedge\star g^{\alpha}\nonumber \\
 & =\star(\theta_{~\gamma}^{\alpha}\wedge g^{\gamma})\wedge g^{\beta}-\star(\theta_{~\gamma}^{\beta}\wedge g^{\gamma})\wedge g^{\alpha}-2\star\theta^{\alpha\beta}\nonumber \\
 & -i^{\gamma}\theta_{\gamma}^{~\alpha}\wedge\star g^{\beta}+i^{\gamma}\theta_{\gamma}^{~\beta}\wedge\star g^{\alpha}\nonumber \\
 & =\delta g^{\beta}\wedge\star g^{\alpha}-\delta g^{\alpha}\wedge\star g^{\beta}+g^{\alpha}\wedge\star dg^{\beta}-g^{\beta}\wedge\star dg^{\alpha}-2\star\theta^{\alpha\beta}\text{.}\label{K1}
\end{align}
But from Eq.(\ref{Levi-Civita}),
\begin{align*}
\theta^{\alpha\beta} & =i^{\beta}dg^{\alpha}-i^{\alpha}dg^{\beta}+\frac{1}{2}i^{\alpha}i^{\beta}\left(g_{\gamma}\wedge dg^{\gamma}\right)\\
 & =i^{\alpha}\star\left(\star dg^{\beta}\right)-i^{\beta}\star\left(\star dg^{\alpha}\right)+\frac{1}{2}i^{\alpha}i^{\beta}\left(g_{\gamma}\wedge dg^{\gamma}\right)\\
 & =\star\left(g^{\alpha}\wedge\star dg^{\beta}\right)-\star\left(g^{\beta}\wedge\star dg^{\alpha}\right)+\frac{1}{2}i^{\alpha}i^{\beta}\left(g_{\gamma}\wedge dg^{\gamma}\right)\text{,}
\end{align*}
so that
\begin{equation}
2\star\theta^{\alpha\beta}=2g^{\alpha}\wedge\star dg^{\beta}-2g^{\beta}\wedge\star dg^{\alpha}+\star i^{\alpha}i^{\beta}\left(g_{\gamma}\wedge dg^{\gamma}\right)\text{.}\label{K2}
\end{equation}
Thus, from Eqs.(\ref{K1}) and (\ref{K2}),
\begin{align*}
d\star\left(g^{\alpha}\wedge g^{\beta}\right) & =\delta g^{\beta}\wedge\star g^{\alpha}-\delta g^{\alpha}\wedge\star g^{\beta}+g^{\alpha}\wedge\star dg^{\beta}-g^{\beta}\wedge\star dg^{\alpha}\\
 & +\star i^{\alpha}i^{\beta}\left(g_{\gamma}\wedge dg^{\gamma}\right)\text{. }\blacksquare
\end{align*}

\subsection{The Non-Metricity Connection\ $\mathfrak{D}$\label{NMCS}}

\begin{definition} A connection $\mathfrak{D}$ in a parallelizable
manifold $M$ is a Non-Metricity \emph{(NM)} connection if and only
if $\mathfrak{D}$ is torsionless and there exists a cobase $\left(g_{\alpha}\right)$
for which the connection $1$--forms $(\omega_{~\beta}^{\alpha})$
of $\mathfrak{D}$ relative to $\left(g_{\alpha}\right)$ satisfy
$\omega_{\left[\alpha\beta\right]}=0$. Or, equivalently, that
\[
\omega_{\alpha\beta}=\mathcal{A}_{\alpha\beta}\text{, for all }0\leq\alpha,\beta<\dim\left(M\right)-1\text{,}
\]
where $(\mathcal{A}_{~\beta}^{\alpha})$ are the non-metricity $1$--forms
of $\mathfrak{D}$ relative to $\left(g_{\alpha}\right)$. Then $\left(g_{\alpha}\right)$
is called an adapted cobase of $\mathfrak{D}$. \end{definition}

\begin{example} Let $M$ be a parallelizable manifold, $\left(x^{\alpha}\right)$
a chart defined on some open neighborhood $U\subset M$ and $g=\eta_{\alpha\beta}dx^{\alpha}\otimes dx^{\beta}$
a metric induced by the cobase field $\left(dx^{\alpha}\right)$.
The Levi-Civita connection of $\left(M,\mathbf{g}\right)$ is a NM
connection, since its connection $1$--forms vanishes. This is the
trivial NM connection. \end{example}

\begin{example} Let $\mathbb{R}^{3}$ be the Euclidean $3$--space,
so that if $\left(r,\theta,\varphi\right)$ are polar coordinates
in $U\subset\mathbb{R}^{3}$, its metric $\mathbf{g}|_{U}=dr\otimes dr+r^{2}\left(d\theta\otimes d\theta+\sin^{2}\theta d\varphi\otimes d\varphi\right)$.
Let $\left(g^{i}\right)_{0\leq i\leq2}$ be the coframe field $g^{0}=dr$,
$g^{1}=rd\theta$ and $g^{3}=r\sin\theta d\varphi$. A NM connection
$\mathfrak{D}$ in $U$ with adapted cobase $\left(g^{i}\right)$
is defined as follows. Let $(\mathcal{A}_{~\beta}^{\alpha})\in$ $%TCIMACRO{\tbigwedge^{1}}%
%BeginExpansion
{\textstyle \bigwedge^{1}}%EndExpansion
M$ be such that,
\[
\mathcal{A}_{~0}^{1}=\frac{1}{r}g^{1}\text{, \ \ }\mathcal{A}_{~2}^{1}=\frac{1}{r\tan\theta}g^{2}\text{, \ \ }\mathcal{A}_{~0}^{2}=\frac{1}{r}g^{2}\text{,}
\]
and, for arbitrary differentiable functions $f,g,h:\mathbb{R}^{3}\longrightarrow\mathbb{R}$,
let $\mathcal{A}_{~0}^{0}=fg^{0}$, $\mathcal{A}_{~1}^{1}=gg^{1}$
and $\mathcal{A}_{~2}^{2}=hg^{2}$. One easily verify that $dg_{i}=-\mathcal{A}_{ij}\wedge g^{j}$
for $0\leq i,j\leq2$. By declaring $(\mathcal{A}_{~\beta}^{\alpha})$
the non-metricity $1$--forms of $\mathfrak{D}$, $\mathfrak{D}$
is completely defined. So our desired NM connection exists but cannot
be unique, given the arbitrarity of the diagonal elements of $(\mathcal{A}_{~\beta}^{\alpha})$.
\end{example}

\begin{example} \label{EXIST}Let $\left(M,\mathbf{g}\right)$ be
a $3$--dimensional Riemannian space and $\left(x,y,z\right)$ an
orthogonal chart on $M$ so that $\mathbf{g}=f^{2}dx\otimes dx+g^{2}dy\otimes dy+h^{2}dz\otimes dz$
for some functions $f,g,h:M\longrightarrow\mathbb{R}$. Let $\left(g^{i}\right)_{0\leq i\leq2}$
be the coframe field $g^{0}=fdx$, $g^{1}=gdy$ and $g^{2}=hdz$.
Define a NM connection $\mathfrak{D}$ with non-metricity $1$--forms
$(\mathcal{A}_{~\beta}^{\alpha})$ as follows. Let,
\[
\mathcal{A}_{~1}^{0}=\frac{f_{y}}{fg}g^{0}+\frac{g_{x}}{fg}g^{1}\text{, \ \ }\mathcal{A}_{~2}^{0}=\frac{f_{z}}{fh}g^{0}+\frac{h_{x}}{fh}g^{2}\text{,}
\]
\[
\mathcal{A}_{~2}^{1}=\frac{g_{z}}{gh}g^{1}+\frac{h_{y}}{gh}g^{2}\text{,}
\]
and define $\mathcal{A}_{~0}^{0}=Fg^{0}$, $\mathcal{A}_{~1}^{1}=Gg^{1}$
and $\mathcal{A}_{~2}^{2}=Hg^{2}$ for any functions $F,G,H:M\longrightarrow\mathbb{R}$.
So it is easy to prove that $dg_{i}=-\mathcal{A}_{ij}\wedge g^{j}$
for $0\leq i,j\leq2$, and then $\mathfrak{D}$ is, up to the diagonal
elements of $(\mathcal{A}_{~\beta}^{\alpha})$, completely defined.
\end{example}

\begin{remark} \label{EXISTENCE}The obvious generalization of the
following example shows that, given any metric on a manifold $M$,
we can always define locally a NM connection with an adapted cobase
field which is a coframe field in $\left(M,\mathbf{g}\right)$. \end{remark}

There exists an identity relating the Levi-Civita connection $1$--forms
relative to an adapted cobase of a NM connection to the components
of the non-metricity $2$--forms of that NM connection. This relation
is expressed in the following Corollary.

\begin{corollary} \label{Non-metricity C.}Let $\mathfrak{D}$ be
a NM connection on a parallelizable manifold $M$ and $\mathbf{Q}_{\alpha\beta\gamma}$
the components of the non-metricity $2$--forms $\left(\mathbf{Q}_{\gamma}\right)$
of $\mathfrak{D}$ relative to an adapted cobase $\left(g_{\alpha}\right)$
of $\mathfrak{D}$\emph{. }The connection $1$--forms $(\theta_{~\beta}^{\alpha})$
of the Levi-Civita connection of the pseudo-Riemannian manifold $\left(M,\mathbf{g}\right)$
\emph{(where }$g=\eta_{\alpha\beta}g^{\alpha}\otimes g^{\beta}$\emph{)}
satisfy
\[
\theta_{\alpha\beta}=\mathbf{Q}_{\alpha\beta\gamma}\wedge g^{\gamma}\text{.}
\]

\end{corollary}

Indeed, by the decomposition formula of Corollary \ref{Decomposition},
\[
\omega_{\left[\alpha\beta\right]}=\omega_{\alpha\beta}-\mathcal{A}_{\alpha\beta}=\theta_{\alpha\beta}-\mathfrak{T}_{\alpha\beta}-\mathbf{Q}_{\alpha\beta\gamma}\wedge g^{\gamma}.
\]
But by the hypothesis on $\mathfrak{D}$, $\omega_{\left[\alpha\beta\right]}=\mathfrak{T}_{\alpha\beta}=0$.
$\blacksquare$

\section{Gravitation and Non-Metricity\label{GNM}}

In section \ref{DFE}, we define and utilize the non-metricity gauge
to formulate a theory of gravitation, whose field equations are derived
from the variational principle. We discuss the form of these field
equations when written in the gravitational Lorenz gauge, and deduce
a force law for the matter currents coupled to the gravitational field.

Then, in section \ref{NM},\ it is shown that the field equations
derived in section \ref{DFE} can be expressed completely in terms
of the components $\mathbf{Q}_{\alpha\beta\gamma}$ of the non-metricity
$2$--forms of Definition \ref{Non-Metricity Field}. We conclude
therefore that the gravitational field may be interpreted as the non-metricity
of a flat torsionless connection. Finally, it is exemplified how the
non-metricity encodes information about the gravitational field in
the Schwarzschild solution.

\subsection{Field Equations\label{DFE}}

The formulation of our theory is based in the following gravitational
Lagrangian density, discovered by Wallner \cite{W} and which is (as
proven in Appendix \ref{EHL}) equivalent to the Einstein-Hilbert
Lagrangian density.

\begin{definition} \label{WL}Let $M$ be a four--dimensional parallelizable
manifold. The Wallner Lagrangian density \emph{(WL)} $\mathcal{L}:(%TCIMACRO{\tbigwedge^{1}}%
%BeginExpansion
{\textstyle \bigwedge^{1}}%EndExpansion
M)^{4}\times(%TCIMACRO{\tbigwedge^{2}}%
%BeginExpansion
{\textstyle \bigwedge^{2}}%EndExpansion
M)^{4}\longrightarrow%TCIMACRO{\tbigwedge^{4}}%
%BeginExpansion
{\textstyle \bigwedge^{4}}%EndExpansion
M$ is
\[
\mathcal{L}=\frac{1}{2}g_{\alpha}\wedge dg^{\beta}\wedge\star\left(g_{\beta}\wedge dg^{\alpha}\right)-\frac{1}{4}g_{\alpha}\wedge dg^{\alpha}\wedge\star\left(g_{\beta}\wedge dg^{\beta}\right)\text{,}
\]
where $\star:%TCIMACRO{\tbigwedge^{p}}%
%BeginExpansion
{\textstyle \bigwedge^{p}}%EndExpansion
M\longrightarrow%TCIMACRO{\tbigwedge^{4-p}}%
%BeginExpansion
{\textstyle \bigwedge^{4-p}}%EndExpansion
M$ is the Hodge dual relative to the metric $\mathbf{g}=\eta_{\alpha\beta}g^{\alpha}\otimes g^{\beta}$
induced by $\left(g_{\alpha}\right)$ and to the orientation $\tau=g^{0}\wedge g^{1}\wedge g^{2}\wedge g^{3}$
of $M$ \cite{Thirring} \cite{GS}. \end{definition}

From now on, cobase fields shall be called \textit{gravitational potentials}.
Our theory begins with the assumption that the gravitational potentials
are cobases adapted to some NM connection $\mathfrak{D}$, and their
set will be denoted by $%TCIMACRO{\tbigwedge}%
%BeginExpansion
{\textstyle \bigwedge}%EndExpansion
\mathfrak{D}$. (By the examples of the last section, we see that $%TCIMACRO{\tbigwedge}%
%BeginExpansion
{\textstyle \bigwedge}%EndExpansion
\mathfrak{D\neq\emptyset}$). This assumption may be referred to as the \textit{non-metricity
gauge}. From Remark \ref{EXISTENCE} and the examples shown above,
one should see that any gravitational field can indeed be represented
by a tetrad in $%TCIMACRO{\tbigwedge}%
%BeginExpansion
{\textstyle \bigwedge}%EndExpansion
\mathfrak{D}$.

\begin{lemma} \label{REST}The restriction $\mathcal{L}|_{\wedge\mathfrak{D}}$
of the WL to cobases belonging to $%TCIMACRO{\tbigwedge}%
%BeginExpansion
{\textstyle \bigwedge}%EndExpansion
\mathfrak{D}$ is
\[
\mathcal{L}|_{\wedge\mathfrak{D}}=\frac{1}{2}g_{\alpha}\wedge dg^{\beta}\wedge\star\left(g_{\beta}\wedge dg^{\alpha}\right)\text{.}
\]

\end{lemma}

Let $\left(g_{\alpha}\right)\in%TCIMACRO{\tbigwedge}%
%BeginExpansion
{\textstyle \bigwedge}%EndExpansion
\mathfrak{D}$. So there exists $(\mathcal{A}_{~\beta}^{\alpha})\in%TCIMACRO{\tbigwedge^{1}}%
%BeginExpansion
{\textstyle \bigwedge^{1}}%EndExpansion
M$ such that $\mathcal{A}_{\alpha\beta}=\mathcal{A}_{\beta\alpha}$
and $dg_{\alpha}=-\mathcal{A}_{\alpha\beta}\wedge g^{\beta}$. Thus
$g_{\alpha}\wedge dg^{\alpha}=g_{\alpha}\wedge g_{\beta}\wedge\mathcal{A}^{\alpha\beta}=0$,
and only the first term of the WL remains. $\blacksquare$

\begin{remark} Any field theory whose Lagrangian can be formulated
in terms of differential forms is invariant under diffeomorphism transformations
\cite{Rodrigues} \cite{Burke}. Therefore, the restriction of the
WL to gravitational potentials in $%TCIMACRO{\tbigwedge}%
%BeginExpansion
{\textstyle \bigwedge}%EndExpansion
\mathfrak{D}$ cannot affect the diffeomorphism invariance of the resulting field
equations. \end{remark}

In the following two Lemmas, we determine the variation of the restricted
WL. In order to avoid confusion with the codifferential operator $\delta$,
lets denote by $\overline{\mathbb{\delta}}$ a variation of $\left(g_{\alpha}\right)$.

Recall that a variation of $\left(g_{\alpha}\right)$ imply in a variation
of $\star$, as the definition of the Hodge dual involves the metric
$\mathbf{g}=\eta_{\alpha\beta}g^{\alpha}\otimes g^{\beta}$. We account
for the Hodge variation by means of the following result.

\begin{lemma} \label{VAR}Let $\omega:%TCIMACRO{\tbigwedge^{1}}%
%BeginExpansion
{\textstyle \bigwedge^{1}}%EndExpansion
M\longrightarrow%TCIMACRO{\tbigwedge^{3}}%
%BeginExpansion
{\textstyle \bigwedge^{3}}%EndExpansion
M$ be a function of $\left(g_{\alpha}\right)$, that is, $\omega=\omega\left(g_{\alpha}\right)$.
Under a variation $\overline{\mathbb{\delta}}$ of $\left(g_{\alpha}\right)$,
\[
\frac{1}{2}\overline{\mathbb{\delta}}\left(\omega\wedge\star\omega\right)=\overline{\mathbb{\delta}}\omega\wedge\star\omega-\overline{\mathbb{\delta}}g_{\alpha}\wedge\left(i^{\alpha}\omega\wedge\star\omega+\frac{1}{2}\star\left(\omega\wedge\star\omega\right)\wedge\star g^{\alpha}\right)\text{.}
\]

\end{lemma}

In fact, first suppose that $\overline{\mathbb{\delta}}\omega=0$.
So the variation of $\omega\wedge\star\omega$ derives exclusively
from that of the Hodge dual: 
\begin{equation}
\overline{\mathbb{\delta}}\left(\omega\wedge\star\omega\right)=\omega\wedge\overline{\mathbb{\delta}}\left(\star\omega\right)\text{.}\label{C1}
\end{equation}
On the other hand, the identity 
\[
g^{\alpha}\wedge g^{\beta}\wedge g^{\gamma}\wedge\star\omega=\omega\wedge\star\left(g^{\alpha}\wedge g^{\beta}\wedge g^{\gamma}\right)
\]
implies that
\begin{equation}
g^{\alpha}\wedge g^{\beta}\wedge g^{\gamma}\wedge\overline{\mathbb{\delta}}\left(\star\omega\right)=\omega\wedge\overline{\mathbb{\delta}}\left(\star\left(g^{\alpha}\wedge g^{\beta}\wedge g^{\gamma}\right)\right)-\overline{\mathbb{\delta}}\left(g^{\alpha}\wedge g^{\beta}\wedge g^{\gamma}\right)\wedge\star\omega\text{.}\label{C1'}
\end{equation}
The first term above can be written as
\[
\overline{\mathbb{\delta}}\left(\star\left(g^{\alpha}\wedge g^{\beta}\wedge g^{\gamma}\right)\right)=\overline{\mathbb{\delta}}g_{\delta}\wedge\varepsilon^{\alpha\beta\gamma\delta}=\overline{\mathbb{\delta}}g_{\delta}\wedge\star\left(g^{\alpha}\wedge g^{\beta}\wedge g^{\gamma}\wedge g^{\delta}\right)\text{,}
\]
while the second as
\[
\overline{\mathbb{\delta}}\left(g^{\alpha}\wedge g^{\beta}\wedge g^{\gamma}\right)=\overline{\mathbb{\delta}}g^{\alpha}\wedge g^{\beta}\wedge g^{\gamma}+g^{\alpha}\wedge\overline{\mathbb{\delta}}g^{\beta}\wedge g^{\gamma}+g^{\alpha}\wedge g^{\beta}\wedge\overline{\mathbb{\delta}}g^{\gamma}\text{.}
\]
Therefore, Eq.(\ref{C1'}) yields
\begin{align}
 & g^{\alpha}\wedge g^{\beta}\wedge g^{\gamma}\wedge\overline{\mathbb{\delta}}\left(\star\omega\right)\nonumber \\
 & =-\overline{\mathbb{\delta}}g_{\delta}\wedge\omega\wedge\star\left(g^{\alpha}\wedge g^{\beta}\wedge g^{\gamma}\wedge g^{\delta}\right)\nonumber \\
 & -\left(\overline{\mathbb{\delta}}g^{\alpha}\wedge g^{\beta}\wedge g^{\gamma}+g^{\alpha}\wedge\overline{\mathbb{\delta}}g^{\beta}\wedge g^{\gamma}+g^{\alpha}\wedge g^{\beta}\wedge\overline{\mathbb{\delta}}g^{\gamma}\right)\wedge\star\omega\text{.}\label{C2}
\end{align}
Now, let $\omega_{\alpha\beta\gamma}=i_{\gamma}i_{\beta}i_{\alpha}\omega$.
Hence, Eqs.(\ref{C1}) and (\ref{C2}) gives
\begin{align}
\overline{\mathbb{\delta}}\left(\omega\wedge\star\omega\right) & =\frac{1}{3!}\omega_{\alpha\beta\gamma}g^{\alpha}\wedge g^{\beta}\wedge g^{\gamma}\wedge\overline{\mathbb{\delta}}\left(\star\omega\right)\nonumber \\
 & =-\overline{\mathbb{\delta}}g_{\alpha}\wedge\omega\wedge\star\left(\omega\wedge g^{\alpha}\right)-\overline{\mathbb{\delta}}g_{\alpha}\wedge\left(\frac{1}{2}\omega_{~\beta\gamma}^{\alpha}g^{\beta}\wedge g^{\gamma}\right)\wedge\star\omega\nonumber \\
 & =-\overline{\mathbb{\delta}}g_{\alpha}\wedge\left(\omega\wedge i^{\alpha}\star\omega+i^{\alpha}\omega\wedge\star\omega\right)\text{.}\label{C3}
\end{align}
Since 
\[
\omega\wedge i^{\alpha}\star\omega=-i^{\alpha}\left(\omega\wedge\star\omega\right)+i^{\alpha}\omega\wedge\star\omega=-\left\langle \omega|\omega\right\rangle \star g^{\alpha}+i^{\alpha}\omega\wedge\star\omega\text{,}
\]
and $-\left\langle \omega|\omega\right\rangle =-i_{\omega}\star\left(\star\omega\right)=-\star\left(\star\omega\wedge\omega\right)=\star\left(\omega\wedge\star\omega\right)$,
Eq.(\ref{C3}) is equivalent to
\[
\frac{1}{2}\overline{\mathbb{\delta}}\left(\omega\wedge\star\omega\right)=-\overline{\mathbb{\delta}}g_{\alpha}\wedge\left(i^{\alpha}\omega\wedge\star\omega+\frac{1}{2}\star\left(\omega\wedge\star\omega\right)\wedge\star g^{\alpha}\right)\text{.}
\]
Lastly, by letting $\omega$ depend on $\left(g_{\alpha}\right)$,
the total variation becomes
\[
\frac{1}{2}\overline{\mathbb{\delta}}\left(\omega\wedge\star\omega\right)=\overline{\mathbb{\delta}}\omega\wedge\star\omega-\overline{\mathbb{\delta}}g_{\alpha}\wedge\left(i^{\alpha}\omega\wedge\star\omega+\frac{1}{2}\star\left(\omega\wedge\star\omega\right)\wedge\star g^{\alpha}\right)\text{. }\blacksquare
\]

\begin{lemma} \label{WLV}The variation of $\mathcal{L}|_{\wedge\mathfrak{D}}$
under a variation $\overline{\mathbb{\delta}}$ of $\left(g_{\alpha}\right)\in%TCIMACRO{\tbigwedge}%
%BeginExpansion
{\textstyle \bigwedge}%EndExpansion
\mathfrak{D}$ is, up to an exact differential,
\[
\overline{\mathbb{\delta}}\left(\mathcal{L}|_{\wedge\mathfrak{D}}\right)=d\left(...\right)+\overline{\mathbb{\delta}}g_{\alpha}\wedge\star G^{\alpha}\text{,}
\]
where $\left(G_{\alpha}\right)\in%TCIMACRO{\tbigwedge^{1}}%
%BeginExpansion
{\textstyle \bigwedge^{1}}%EndExpansion
M$ are the Einstein $1$--forms
\begin{align}
\star G_{\alpha} & =dg^{\beta}\wedge\star\left(g_{\beta}\wedge dg_{\alpha}\right)+d\left(g_{\beta}\wedge\star\left(g_{\alpha}\wedge dg^{\beta}\right)\right)-\star\left(\mathcal{L}|_{\wedge\mathfrak{D}}\right)\wedge\star g_{\alpha}\nonumber \\
 & -i_{\alpha}\left(g_{\beta}\wedge dg^{\gamma}\right)\wedge\star\left(g_{\gamma}\wedge dg^{\beta}\right)\text{.}\label{EINST}
\end{align}

\end{lemma}

Indeed, from Lemmas \ref{REST} and \ref{VAR}, 
\begin{align}
\overline{\mathbb{\delta}}\left(\mathcal{L}|_{\wedge\mathfrak{D}}\right) & =\frac{1}{2}\overline{\mathbb{\delta}}\left(g_{\alpha}\wedge dg^{\beta}\wedge\star\left(g_{\beta}\wedge dg^{\alpha}\right)\right)\nonumber \\
 & =\overline{\mathbb{\delta}}\left(g_{\alpha}\wedge dg^{\beta}\right)\wedge\star\left(g_{\beta}\wedge dg^{\alpha}\right)\nonumber \\
 & -\overline{\mathbb{\delta}}g_{\alpha}\wedge\left(i^{\alpha}\left(g_{\beta}\wedge dg^{\gamma}\right)\wedge\star\left(g_{\gamma}\wedge dg^{\beta}\right)+\star\left(\mathcal{L}|_{\wedge\mathfrak{D}}\right)\wedge\star g^{\alpha}\right)\text{.}\label{G1}
\end{align}
But the first term gives
\begin{align}
 & \overline{\mathbb{\delta}}\left(g_{\alpha}\wedge dg^{\beta}\right)\wedge\star\left(g_{\beta}\wedge dg^{\alpha}\right)\nonumber \\
 & =\overline{\mathbb{\delta}}g_{\alpha}\wedge dg^{\beta}\wedge\star\left(g_{\beta}\wedge dg^{\alpha}\right)+g_{\alpha}\wedge d\overline{\mathbb{\delta}}g^{\beta}\wedge\star\left(g_{\beta}\wedge dg^{\alpha}\right)\nonumber \\
 & =\overline{\mathbb{\delta}}g_{\alpha}\wedge dg^{\beta}\wedge\star\left(g_{\beta}\wedge dg^{\alpha}\right)+d\left(...\right)+\overline{\mathbb{\delta}}g_{\alpha}\wedge d\left(g_{\beta}\wedge\star\left(g^{\alpha}\wedge dg^{\beta}\right)\right)\nonumber \\
 & =d\left(...\right)+\overline{\mathbb{\delta}}g_{\alpha}\wedge\left(dg^{\beta}\wedge\star\left(g_{\beta}\wedge dg^{\alpha}\right)+d\left(g_{\beta}\wedge\star\left(g^{\alpha}\wedge dg^{\beta}\right)\right)\right)\text{.}\label{G2}
\end{align}
Then, from Eqs.(\ref{G1}) and (\ref{G2}),
\begin{align*}
\overline{\mathbb{\delta}}\left(\mathcal{L}|_{\wedge\mathfrak{D}}\right) & =d\left(...\right)\\
 & +\overline{\mathbb{\delta}}g_{\alpha}\wedge\lbrack dg^{\beta}\wedge\star\left(g_{\beta}\wedge dg^{\alpha}\right)+d\left(g_{\beta}\wedge\star\left(g^{\alpha}\wedge dg^{\beta}\right)\right)\\
 & -i^{\alpha}\left(g_{\beta}\wedge dg^{\gamma}\right)\wedge\star\left(g_{\gamma}\wedge dg^{\beta}\right)-\star\left(\mathcal{L}|_{\wedge\mathfrak{D}}\right)\wedge\star g^{\alpha}]\text{.}
\end{align*}
The result follows by recognizing the terms between the square brackets
as $\star G^{\alpha}$ (see Eq.(\ref{EINST})). $\blacksquare$

\bigskip{}

In the next Lemma, we decompose the Einstein $1$--forms in the differential
term $\delta dg_{\alpha}$ together with the gravitational energy-momentum
currents $\mathcal{T}_{\alpha}$ (Eq.(\ref{GEM})). In this way, the
Einstein equation shall assume the form of four coupled equations
with the same structure as the inhomogeneous Maxwell equation, from
which we may derive a simple conservation law (Remark \ref{RODRIGUES}
and Proposition \ref{FIELD EQ}).

\begin{lemma} \label{EINST1-FORM}The Einstein $1$--forms $\left(G_{\alpha}\right)\in%TCIMACRO{\tbigwedge^{1}}%
%BeginExpansion
{\textstyle \bigwedge^{1}}%EndExpansion
M$ of the gravitational potentials $\left(g_{\alpha}\right)$ \emph{(Eq.(\ref{EINST}))}
can be written as
\begin{equation}
G_{\alpha}=-\delta dg_{\alpha}+\mathcal{T}_{\alpha}\text{,}\label{EINSTD}
\end{equation}
where $\left(\mathcal{T}_{\alpha}\right)\in%TCIMACRO{\tbigwedge^{1}}%
%BeginExpansion
{\textstyle \bigwedge^{1}}%EndExpansion
M$ are the gravitational energy-momentum currents:
\begin{align}
\mathcal{T}_{\alpha} & =\frac{1}{2}\star\left(dg_{\beta}\wedge i_{\alpha}\star dg^{\beta}-i_{\alpha}dg_{\beta}\wedge\star dg^{\beta}\right)\nonumber \\
 & +\frac{1}{2}\delta g_{\beta}\wedge\delta g^{\beta}\wedge g_{\alpha}+i_{\beta}d\delta g^{\beta}\wedge g_{\alpha}-i_{\alpha}d\delta g^{\beta}\wedge g_{\beta}\text{.}\label{GEM}
\end{align}

\end{lemma}

In fact, the first term of Eq.(\ref{EINST}) yields
\begin{align}
dg^{\beta}\wedge\star\left(g_{\beta}\wedge dg_{\alpha}\right) & =dg^{\beta}\wedge i_{\beta}\star dg_{\alpha}\nonumber \\
 & =i_{\beta}\left(dg^{\beta}\wedge\star dg_{\alpha}\right)-i_{\beta}dg^{\beta}\wedge\star dg_{\alpha}\nonumber \\
 & =\left\langle dg_{\alpha}|dg^{\beta}\right\rangle \star g_{\beta}+\delta g_{\beta}\wedge g^{\beta}\wedge\star dg_{\alpha}\text{,}\label{H1}
\end{align}
where in the last line we used Lemma \ref{Trick}.

Now, the second term of Eq.(\ref{EINST}) gives
\begin{align*}
d\left(g_{\beta}\wedge\star\left(g_{\alpha}\wedge dg^{\beta}\right)\right) & =d\left(g_{\beta}\wedge i_{\alpha}\star dg^{\beta}\right)\\
 & =d\left(-i_{\alpha}\left(g_{\beta}\wedge\star dg^{\beta}\right)+\star dg_{\alpha}\right)\\
 & =d\star dg_{\alpha}-di_{\alpha}\left(g_{\beta}\wedge\star dg^{\beta}\right)\text{.}
\end{align*}
On the other hand,
\begin{align*}
g_{\beta}\wedge\star dg^{\beta} & =\star^{2}\left(\star dg^{\beta}\wedge g_{\beta}\right)=\star i_{\beta}\star^{2}dg^{\beta}\\
 & =-\star i_{\beta}dg^{\beta}=\delta g_{\beta}\wedge\star g^{\beta}\text{,}
\end{align*}
from Lemma \ref{Trick} again. So,
\begin{align*}
d\left(g_{\beta}\wedge\star\left(g_{\alpha}\wedge dg^{\beta}\right)\right) & =\star^{2}\left(d\star dg_{\alpha}\right)+d\left(\delta g_{\beta}\wedge\star\left(g_{\alpha}\wedge g^{\beta}\right)\right)\\
 & =-\star\delta g_{\alpha}+d\delta g_{\beta}\wedge\star\left(g_{\alpha}\wedge g^{\beta}\right)+\delta g_{\beta}\wedge d\star\left(g_{\alpha}\wedge g^{\beta}\right)\text{.}
\end{align*}
Therefore, by applying Lemma \ref{Trick_},
\begin{align}
d\left(g_{\beta}\wedge\star\left(g_{\alpha}\wedge dg^{\beta}\right)\right) & =-\star\delta g_{\alpha}+d\delta g_{\beta}\wedge\star\left(g_{\alpha}\wedge g^{\beta}\right)\nonumber \\
 & +\delta g_{\beta}\wedge(\delta g^{\beta}\wedge\star g^{\alpha}-\delta g^{\alpha}\wedge\star g^{\beta}\nonumber \\
 & +g^{\beta}\wedge\star dg^{\alpha}-g^{\alpha}\wedge\star dg^{\beta})\text{.}\label{H2}
\end{align}
where we used that $dg_{\gamma}\wedge g^{\gamma}=0$ in the non-metricity
gauge.

On the other hand, from Lemma \ref{RdS} (proved in Appendix \ref{EHL}),
\begin{align*}
\mathcal{L}|_{\wedge\mathfrak{D}} & =\frac{1}{2}dg_{\beta}\wedge\star dg^{\beta}-\frac{1}{2}\delta g_{\beta}\wedge\star\delta g^{\beta}\\
 & =\left(\frac{1}{2}\left\langle dg_{\beta}|dg^{\beta}\right\rangle -\frac{1}{2}\delta g_{\beta}\wedge\delta g^{\beta}\right)\star1
\end{align*}
Hence
\[
-\star\left(\mathcal{L}|_{\wedge\mathfrak{D}}\right)=\frac{1}{2}\left\langle dg_{\beta}|dg^{\beta}\right\rangle -\frac{1}{2}\delta g_{\beta}\wedge\delta g^{\beta}\text{,}
\]
and the third term of Eq.(\ref{EINST}) becomes
\begin{equation}
-\star\left(\mathcal{L}|_{\wedge\mathfrak{D}}\right)\wedge\star g_{\alpha}=\left(\frac{1}{2}\left\langle dg_{\beta}|dg^{\beta}\right\rangle -\frac{1}{2}\delta g_{\beta}\wedge\delta g^{\beta}\right)\wedge\star g_{\alpha}\text{.}\label{H3}
\end{equation}

Finally, we work on the last term of Eq.(\ref{EINST}), for which
\begin{align}
 & i_{\alpha}\left(g_{\beta}\wedge dg^{\gamma}\right)\wedge\star\left(g_{\gamma}\wedge dg^{\beta}\right)\nonumber \\
 & =dg^{\gamma}\wedge i_{\gamma}\star dg_{\alpha}+i_{\alpha}dg^{\gamma}\wedge g_{\beta}\wedge i_{\gamma}\star dg^{\beta}\text{.}\label{H'4}
\end{align}
First,
\begin{align}
dg^{\gamma}\wedge i_{\gamma}\star dg_{\alpha} & =i_{\gamma}\left(dg^{\gamma}\wedge\star dg_{\alpha}\right)-i_{\gamma}dg^{\gamma}\wedge\star dg_{\alpha}\nonumber \\
 & =\left\langle dg_{\alpha}|dg^{\gamma}\right\rangle \star g_{\gamma}+\delta g_{\gamma}\wedge g^{\gamma}\wedge\star dg_{\alpha}\text{.}\label{H4}
\end{align}
Second,
\begin{align}
 & i_{\alpha}dg^{\gamma}\wedge g_{\beta}\wedge i_{\gamma}\star dg^{\beta}\nonumber \\
 & =i_{\gamma}\left(i_{\alpha}dg^{\gamma}\wedge g_{\beta}\wedge\star dg^{\beta}\right)-i_{\gamma}\left(i_{\alpha}dg^{\gamma}\wedge g_{\beta}\right)\wedge\star dg^{\beta}\nonumber \\
 & =i_{\gamma}\left(i_{\alpha}dg^{\gamma}\wedge\star\star\left(\star dg^{\beta}\wedge g_{\beta}\right)\right)+i_{\alpha}i_{\gamma}dg^{\gamma}\wedge g_{\beta}\wedge\star dg^{\beta}+i_{\alpha}dg^{\gamma}\wedge\eta_{\beta\gamma}\star dg^{\beta}\nonumber \\
 & =-i_{\gamma}\left(i_{\alpha}dg^{\gamma}\wedge\star i_{\beta}dg^{\beta}\right)+i_{\alpha}i_{\gamma}dg^{\gamma}\wedge g_{\beta}\wedge\star dg^{\beta}+i_{\alpha}dg_{\beta}\wedge\star dg^{\beta}\nonumber \\
 & =\delta g_{\beta}\wedge i_{\gamma}\left(i_{\alpha}dg^{\gamma}\wedge\star g^{\beta}\right)-\delta g_{\alpha}\wedge\star\star\left(\star dg^{\beta}\wedge g_{\beta}\right)+\left\langle dg_{\beta}|dg^{\beta}\right\rangle \star g_{\alpha}\nonumber \\
 & -dg_{\beta}\wedge i_{\alpha}\star dg^{\beta}\nonumber \\
 & =\delta g_{\beta}\wedge i_{\gamma}\left(i_{\alpha}dg^{\gamma}\wedge\star g^{\beta}\right)-\delta g_{\alpha}\wedge\delta g_{\beta}\wedge\star g^{\beta}+\left\langle dg_{\beta}|dg^{\beta}\right\rangle \star g_{\alpha}\nonumber \\
 & -dg_{\beta}\wedge i_{\alpha}\star dg^{\beta}\text{.}\label{H5}
\end{align}
Third,
\begin{align}
\delta g_{\beta}\wedge i_{\gamma}\left(i_{\alpha}dg^{\gamma}\wedge\star g^{\beta}\right) & =\delta g_{\beta}\wedge i_{\gamma}\left(g^{\beta}\wedge\star i_{\alpha}dg^{\gamma}\right)\nonumber \\
 & =\delta g_{\beta}\wedge\star i_{\alpha}dg^{\beta}-\delta g_{\beta}\wedge g^{\beta}\wedge i_{\gamma}\star i_{\alpha}dg^{\gamma}\nonumber \\
 & =-\delta g_{\beta}\wedge\star i_{\alpha}\star\left(\star dg^{\beta}\right)+\delta g_{\beta}\wedge g^{\beta}\wedge i_{\gamma}\star i_{\alpha}\star\left(\star dg^{\gamma}\right)\nonumber \\
 & =-\delta g_{\beta}\wedge\star dg^{\beta}\wedge g_{\alpha}+\delta g_{\beta}\wedge g^{\beta}\wedge i_{\gamma}\left(\star dg^{\gamma}\wedge g_{\alpha}\right)\nonumber \\
 & =-\delta g_{\beta}\wedge\star dg^{\beta}\wedge g_{\alpha}+\delta g_{\beta}\wedge\star dg_{\alpha}\wedge g^{\beta}\text{,}\label{H6}
\end{align}
where the non-metricity gauge has been used again in $i_{\gamma}\star dg^{\gamma}=0$.
Substituting Eq.(\ref{H6}) in Eq.(\ref{H5}), we obtain
\begin{align}
 & i_{\alpha}dg^{\gamma}\wedge g_{\beta}\wedge i_{\gamma}\star dg^{\beta}\nonumber \\
 & =\left\langle dg_{\beta}|dg^{\beta}\right\rangle \star g_{\alpha}-dg_{\beta}\wedge i_{\alpha}\star dg^{\beta}\nonumber \\
 & +\delta g_{\beta}\wedge(\star dg_{\alpha}\wedge g^{\beta}-\star dg^{\beta}\wedge g_{\alpha}-\delta g_{\alpha}\wedge\star g^{\beta})\text{.}\label{H7}
\end{align}
Finally, from Eqs.(\ref{H'4}), (\ref{H4}) and (\ref{H7}),
\begin{align}
 & i_{\alpha}\left(g_{\beta}\wedge dg^{\gamma}\right)\wedge\star\left(g_{\gamma}\wedge dg^{\beta}\right)\nonumber \\
 & =\left\langle dg_{\beta}|dg^{\beta}\right\rangle \star g_{\alpha}+\left\langle dg_{\alpha}|dg^{\gamma}\right\rangle \star g_{\gamma}-dg_{\beta}\wedge i_{\alpha}\star dg^{\beta}\nonumber \\
 & +\delta g_{\beta}\wedge(2\star dg_{\alpha}\wedge g^{\beta}-\star dg^{\beta}\wedge g_{\alpha}-\delta g_{\alpha}\wedge\star g^{\beta})\text{.}\label{H8}
\end{align}

Lastly, substituting Eqs.(\ref{H1}), (\ref{H2}), (\ref{H3}) and
(\ref{H8}) in Eq.(\ref{EINST}), performing some algebraic cancelling
and reorganizing the terms, we obtain
\begin{align*}
\star G_{\alpha} & =-\star d\delta g_{\alpha}+\left(dg_{\beta}\wedge i_{\alpha}\star dg^{\beta}-\frac{1}{2}\left\langle dg_{\beta}|dg^{\beta}\right\rangle \star g_{\alpha}\right)+\frac{1}{2}\delta g_{\beta}\wedge\delta g^{\beta}\wedge\star g_{\alpha}\\
 & +d\delta g_{\beta}\wedge\star\left(g_{\alpha}\wedge g^{\beta}\right)\text{,}
\end{align*}
which we recognize as being equivalent to Eq.(\ref{EINSTD}). $\blacksquare$

\begin{remark} \label{ELECTROMAG}The first term of \emph{Eq.(\ref{GEM})
}for the gravitational energy-momentum current $\mathcal{T}_{\alpha}$
posses the same structure\footnote{Up to a minus sign.}\emph{ }as
the energy-momentum currents of the electromagnetic field $\mathbf{F}\in%TCIMACRO{\tbigwedge^{2}}%
%BeginExpansion
{\textstyle \bigwedge^{2}}%EndExpansion
M$, namely \cite{Thirring} \cite{GS}
\[
\mathrm{T}_{\alpha}=-\frac{1}{2}\star\left(\mathbf{F}\wedge i_{\alpha}\star\mathbf{F}-i_{\alpha}\mathbf{F}\wedge\star\mathbf{F}\right)\text{.}
\]
The following terms of $\mathcal{T}_{\alpha}$ consists of contributions
derived from the codifferential $\delta g_{\alpha}$, all of which
vanishes in the gravitational Lorenz gauge, yielding a simple set
of equations for the gravitational field \emph{(see Remark \ref{LORENZ})}.
\end{remark}

\begin{remark} \label{TW}The expression for the Einstein $1$--forms
given in Lemma \ref{EINST1-FORM} can be derived from a decomposition
\emph{(due to Thirring and Wallner \cite{W} \cite{TW} and rediscovered
by Sparling \cite{S})} in which
\begin{equation}
\star G_{\alpha}=d\star\pi_{\alpha}+\star t_{\alpha}\text{,}\label{WALLNER-THIRRING}
\end{equation}
where $\left(\pi_{\alpha}\right)\in%TCIMACRO{\tbigwedge^{2}}%
%BeginExpansion
{\textstyle \bigwedge^{2}}%EndExpansion
M$ are known as the superpotentials and $\left(t_{\alpha}\right)\in%TCIMACRO{\tbigwedge^{1}}%
%BeginExpansion
{\textstyle \bigwedge^{1}}%EndExpansion
M$ as the pseudo-currents. For the record, $\pi_{\alpha}$ and $t_{\alpha}$
are given in terms of the Levi-Civita connection $1$--forms $(\theta_{~\beta}^{\alpha})$
by
\[
\star\pi_{\gamma}=\frac{1}{2}\theta_{\alpha\beta}\wedge\star\left(g^{\alpha}\wedge g^{\beta}\wedge g_{\gamma}\right)\text{,}
\]
\[
\star t_{\gamma}=-\frac{1}{2}\theta_{\alpha\beta}\wedge(\theta_{\gamma\delta}\wedge\star\left(g^{\alpha}\wedge g^{\beta}\wedge g^{\delta}\right)+\theta_{~\delta}^{\beta}\wedge\star\left(g^{\alpha}\wedge g^{\delta}\wedge g_{\gamma}\right))\text{,}
\]
both of which can be derived from the formula of $G_{\alpha}$ in
terms of the curvature $2$--forms $(\mathcal{R}_{~\beta}^{\alpha})$
of the Levi-Civita connection of $\left(M,\mathbf{g}\right)$, i.e.
\cite{Thirring} \cite{GS}
\[
G_{\alpha}=\frac{1}{2}\mathcal{R}_{\beta\gamma}\wedge\star\left(g^{\beta}\wedge g^{\gamma}\wedge g_{\alpha}\right)\text{,}
\]
and from the second Cartan structural equation. By employing \emph{Eq.(\ref{Levi-Civita})},
it is possible to express the Thirring-Wallner form \emph{(Eq.(\ref{WALLNER-THIRRING}))
}just in terms of $g_{\alpha}$ and $dg_{\alpha}$, a fact which was
one of the motivations leading Wallner to the Lagrangian of Definition
\ref{WL} \emph{\cite{W}}. \end{remark}

\begin{remark} \label{RODRIGUES}On the other hand, the idea of expressing
the Einstein $1$--forms as 
\[
G_{\alpha}=-\delta dg_{\alpha}+\mathcal{T}_{\alpha}\text{,}
\]
so that the Einstein equations assumes the formal structure of four
inhomogeneous Maxwell equations $\delta dg_{\alpha}=\mathcal{T}_{\alpha}+\mathcal{J}_{\alpha}$
coupled to the matter energy-momentum currents $\left(\mathcal{J}_{\alpha}\right)$,
is due to Rodrigues \cite{R}. From this, we deduce the conservation
law 
\[
\mathcal{\delta}\left(\mathcal{T}_{\alpha}+\mathcal{J}_{\alpha}\right)=0\text{,}
\]
which suggests the identification of $\left(\mathcal{T}_{\alpha}\right)$
as the energy-momentum currents of the gravitational field, instead
of the archaic energy-momentum pseudo-tensors usually found in the
literature \cite{Anderson} \cite{Landau}. However, in virtue of
the non-metricity gauge adopted here, our \emph{Eq.(\ref{GEM}) }for
the gravitational energy-momentum currents is simpler and even more
appealing physically than the one appearing in \cite{R}. \end{remark}

In the following paragraphs, the variational principle will be finally
applied to derive the gravitational field equations. Then we show
in Example \ref{LORENZ} that, if written in the Lorenz gauge, our
field equations becomes a system of coupled Proca equations for the
gravitational potentials, whose application to the study of gravitational
and electromagnetic waves is briefly outlined in Examples \ref{WAVES 1}
and \ref{WAVES 2}.

\bigskip{}

To account for the matter energy-momentum currents $\left(\mathcal{J}_{\alpha}\right)\in%TCIMACRO{\tbigwedge^{1}}%
%BeginExpansion
{\textstyle \bigwedge^{1}}%EndExpansion
M$ (and possibly other classical fields), let $\mathcal{L}_{m}:(%TCIMACRO{\tbigwedge^{1}}%
%BeginExpansion
{\textstyle \bigwedge^{1}}%EndExpansion
M)^{4}\times(%TCIMACRO{\tbigwedge^{2}}%
%BeginExpansion
{\textstyle \bigwedge^{2}}%EndExpansion
M)^{4}\longrightarrow%TCIMACRO{\tbigwedge^{4}}%
%BeginExpansion
{\textstyle \bigwedge^{4}}%EndExpansion
M$ represent the matter Lagrangian. On what follows, it is supposed
that under a variation $\overline{\delta}$ of $\left(g_{\alpha}\right)\in%TCIMACRO{\tbigwedge}%
%BeginExpansion
{\textstyle \bigwedge}%EndExpansion
\mathfrak{D}$, 
\begin{equation}
\overline{\delta}\mathcal{L}_{m}=\overline{\delta}g_{\alpha}\wedge\star\mathcal{J}^{\alpha}\text{.}\label{MATTER}
\end{equation}
Details regarding $\mathcal{L}_{m}$ and some matter models are described
in \cite{Thirring} and \cite{GS}.

\begin{proposition} \label{FIELD EQ}Let $\mathcal{S}:%TCIMACRO{\tbigwedge}%
%BeginExpansion
{\textstyle \bigwedge}%EndExpansion
\mathfrak{D}\longrightarrow\mathbb{R}$ be the action functional
\[
\mathcal{S}\left[\left(g_{\alpha}\right)\right]=\int_{\mathcal{V}}\left(\mathcal{L}|_{\wedge\mathfrak{D}}+\mathcal{L}_{m}\right)\left(\left(g_{\alpha}\right),\left(dg_{\alpha}\right)\right)\text{,}
\]
for $\mathcal{V}\subset M$ a compact four--dimensional submanifold.
Thus $\mathcal{S}$ is stationary under a variation $\overline{\delta}$
of $\left(g_{\alpha}\right)\in%TCIMACRO{\tbigwedge}%
%BeginExpansion
{\textstyle \bigwedge}%EndExpansion
\mathfrak{D}$, that is,
\[
\overline{\delta}\mathcal{S}\left[\left(g_{\alpha}\right)\right]=0\text{,}
\]
if and only if the Einstein equations
\begin{equation}
\delta dg_{\alpha}=\mathcal{T}_{\alpha}+\mathcal{J}_{\alpha}\text{,}\label{EINSTEIN EQ}
\end{equation}
are satisfied. \end{proposition}

In fact, one can prove that \cite{Burke} 
\[
\overline{\delta}\mathcal{S}=\overline{\delta}\int_{\mathcal{V}}\left(\mathcal{L}|_{\wedge\mathfrak{D}}+\mathcal{L}_{m}\right)=\int_{\mathcal{V}}\overline{\delta}\left(\mathcal{L}|_{\wedge\mathfrak{D}}+\mathcal{L}_{m}\right)\text{.}
\]
Hence, by Lemma \ref{WLV} and Eq.(\ref{MATTER}), 
\[
\overline{\delta}\mathcal{S=}\int_{\mathcal{V}}\overline{\delta}g_{\alpha}\wedge\star\left(G^{\alpha}+\mathcal{J}^{\alpha}\right)=0\text{,}
\]
so that $G^{\alpha}+\mathcal{J}^{\alpha}=0$, and the Einstein equations
(Eq.(\ref{EINSTEIN EQ})) follows from Lemma \ref{EINST1-FORM}. $\blacksquare$

\begin{remark} The interaction of the gravitational energy-momentum
currents $\left(\mathcal{T}_{\alpha}\right)$ with the gravitational
field is hidden in the old-fashioned geometrical formulation of GR
in the pseudo-Riemannian space $\left(M,\mathbf{g,}\nabla\right)$.
Recall that the Einstein equations are given in terms of the Ricci
$1$--forms $\left(\mathcal{R}_{\alpha}\right)$ of $\left(M,\mathbf{g,}\nabla\right)$
by
\[
\mathcal{R}_{\alpha}-\frac{1}{2}\mathcal{R}g_{\alpha}=\mathcal{J}_{\alpha}\text{.}
\]
In this form, only the matter energy-momentum currents $\left(\mathcal{J}_{\alpha}\right)$
appears in the right-hand side, while the gravitational currents $\left(\mathcal{T}_{\alpha}\right)$
are disguised in the ``geometrical\textquotedblright \ left-hand
side. On the other hand, according to the gravitational equations
which were derived above, from where we read
\[
\delta dg_{\alpha}=\mathcal{T}_{\alpha}+\mathcal{J}_{\alpha}\text{,}
\]
it is seen that both the matter and gravitational energy-momentum
currents contributes to and interacts with the gravitational field,
realizing the physical idea that the gravitational field interacts
with itself. \end{remark}

\begin{example} {[}Lorenz Gauge{]}\label{LORENZ}Lets suppose that
$\left(g_{\alpha}\right)$ satisfy the gravitational Lorenz gauge,
for which
\[
\delta g_{\alpha}=0\text{, \ \ }0\leq\alpha\leq3\text{.}
\]
It follows that the gravitational energy-momentum currents $\left(\mathcal{T}_{\alpha}\right)$
\emph{(recall Eq.(\ref{GEM}))} becomes
\[
\mathcal{T}_{\alpha}=\frac{1}{2}\star\left(dg_{\beta}\wedge i_{\alpha}\star dg^{\beta}-i_{\alpha}dg_{\beta}\wedge\star dg^{\beta}\right)\text{,}
\]
possessing therefore the same structure as the electromagnetic energy-momentum
currents \emph{(compare with Remark \ref{LORENTZ})}. But since 
\begin{align}
i_{\alpha}dg_{\beta}\wedge\star dg^{\beta} & =i_{\alpha}\left(dg_{\beta}\wedge\star dg^{\beta}\right)-dg_{\beta}\wedge i_{\alpha}\star dg^{\beta}\nonumber \\
 & =\left\langle dg_{\beta}|dg^{\beta}\right\rangle \star g_{\alpha}-dg_{\beta}\wedge\star\left(g_{\alpha}\wedge dg^{\beta}\right)\text{,}\label{D0}
\end{align}
the currents $\left(\mathcal{T}_{\alpha}\right)$ can be rewritten
as
\begin{equation}
\mathcal{T}_{\alpha}=i_{dg_{\beta}}\left(g_{\alpha}\wedge dg^{\beta}\right)-\frac{1}{2}\left\langle dg_{\beta}|dg^{\beta}\right\rangle g_{\alpha}\text{.}\label{D1}
\end{equation}
On the other hand, recall that the Laplace-Beltrami operator is defined
by $\square=d\delta+\delta d$ \emph{(see \cite{Thirring} or \cite{GS})},
so that in the Lorenz gauge,
\begin{equation}
\square g_{\alpha}=\left(d\delta+\delta d\right)g_{\alpha}=\delta dg_{\alpha}\text{.}\label{D2}
\end{equation}
Finally, from Eqs.(\ref{D1}) and (\ref{D2}) together with the Einstein
equation \emph{(Eq.(\ref{EINSTEIN EQ}))}, our gravitational field
equations becomes the following system of four coupled Proca equations,
\begin{equation}
\square g_{\alpha}+\frac{1}{2}\left\langle dg_{\beta}|dg^{\beta}\right\rangle g_{\alpha}=i_{dg_{\beta}}\left(g_{\alpha}\wedge dg^{\beta}\right)+\mathcal{J}_{\alpha}\text{,}\label{PROCA EQ}
\end{equation}
whose variable mass $\mathfrak{M}=\frac{1}{2}\left\langle dg_{\beta}|dg^{\beta}\right\rangle $
derives entirely from the gravitational field, while the sources are
both from gravitational and matter origins, namely, $i_{dg_{\beta}}\left(g_{\alpha}\wedge dg^{\beta}\right)$
and $\mathcal{J}_{\alpha}$ respectively. \end{example}

\begin{remark} The variable mass $\mathfrak{M}$ appearing in the
Proca equations for the gravitational potentials in the Lorenz gauge
\emph{(Eq.(\ref{PROCA EQ})) }cannot be identified as the graviton
mass. In fact, if the mass term $\frac{1}{4}m^{2}g_{\alpha}\wedge\star g^{\alpha}$
is included in the WL, it can be proven that the currents appearing
in our field equations \emph{(Eq.(\ref{EINSTEIN EQ}))} receives an
additional mass term, becoming
\begin{equation}
\delta dg_{\alpha}=\mathcal{T}_{\alpha}+m^{2}g_{\alpha}+\mathcal{J}_{\alpha}\text{.}\label{E1}
\end{equation}
This, in turn, adds a correction to the variable mass $\mathfrak{M}$
of our Proca equations, which now reads
\[
\square g_{\alpha}+\left(\frac{1}{2}\left\langle dg_{\beta}|dg^{\beta}\right\rangle +m^{2}\right)g_{\alpha}=i_{dg_{\beta}}\left(g_{\alpha}\wedge dg^{\beta}\right)+\mathcal{J}_{\alpha}\text{.}
\]
Alternatively, if we look at the geometrical formulation of GR in
the pseudo-Riemannian space $\left(M,\mathbf{g,}\nabla\right)$, the
field equations after the inclusion of the mass term $\frac{1}{4}m^{2}g_{\alpha}\wedge\star g^{\alpha}$
are given by
\[
\mathcal{R}_{\alpha}-\frac{1}{2}\mathcal{R}g_{\alpha}+m^{2}g_{\alpha}=\mathcal{J}_{\alpha}\text{.}
\]
The presence of a mass term in the WL, therefore, is equivalent to
the introduction of a cosmological constant. So, while the variable
mass $\mathfrak{M}$ of \emph{Eq.(\ref{PROCA EQ})} is effective in
character, depending on the configuration of the gravitational field,
the mass $m$ \emph{(or, equivalently, a cosmological constant)} interacts
with the gravitational potential by means of a contribution to the
gravitational energy-momentum currents, as seen in \emph{Eq.(\ref{E1}).}
\end{remark}

\begin{example} \label{WAVES 1}It is interesting to note that our
vacuum \emph{(}$\mathcal{J}_{\alpha}=0$\emph{)} gravitational equations
in the Lorenz gauge \emph{(Eq.(\ref{PROCA EQ}))} are given by
\[
\square g_{\alpha}+\frac{1}{2}\left\langle dg_{\beta}|dg^{\beta}\right\rangle g_{\alpha}=i_{dg_{\beta}}\left(g_{\alpha}\wedge dg^{\beta}\right)\text{.}
\]
Therefore, even in the absence of matter, the Proca equations describing
the propagation of the gravitational potentials posses a source, but
which is now purely gravitational in origin. This is a statement of
the non-linearity inherent to the propagation of the gravitational
field, and a mathematical realization of the physical picture that
gravitational disturbances may itself be a source of gravitational
fields. \end{example}

\begin{example} \label{WAVES 2}In electrovacuum, the only contribution
to the ``matter\textquotedblright \ energy-momentum currents $\left(\mathcal{J}_{\alpha}\right)$
derives from the electromagnetic field, for which \emph{(recall Remark
\ref{ELECTROMAG})}
\[
\mathcal{J}_{\alpha}=\frac{1}{2}\star\left(i_{\alpha}\mathbf{F}\wedge\star\mathbf{F-F}\wedge i_{\alpha}\star\mathbf{F}\right)=\frac{1}{2}\left\langle \mathbf{F|F}\right\rangle g_{\alpha}-i_{\mathbf{F}}\left(\mathbf{F\wedge}g_{\alpha}\right)\text{.}
\]
Accordingly, our gravitational field equations in the Lorenz gauge
\emph{(Eq.(\ref{PROCA EQ})) }becomes
\[
\square g_{\alpha}+\frac{1}{2}\left(\left\langle dg_{\beta}|dg^{\beta}\right\rangle -\left\langle \mathbf{F|F}\right\rangle \right)g_{\alpha}=i_{dg_{\beta}}\left(g_{\alpha}\wedge dg^{\beta}\right)-i_{\mathbf{F}}\left(\mathbf{F\wedge}g_{\alpha}\right)\text{.}
\]
We then see that the existence of a nonvanishing electromagnetic field
changes the effective mass of propagation of the gravitational potentials
and introduces a contribution to the source term in the above Proca
equations. This result suggests that the existence of electromagnetic
oscillations may be a source for gravitational pertubation, which
is consonant with many solutions of the Einstein-Maxwell equations
\emph{(and its linearized version)} which describes a coupled system
of gravitational-electromagnetic waves \emph{(some of which are \cite{Bramson},
\cite{Vaidya} and \cite{Roy})}. \end{example}

Our form of the Einstein equation admits a very simple ``force law\textquotedblright ,
which is just the analog of the Newtonian theorem of work and energy
variation.

\begin{definition} \label{FL}Let $\xi\in\sec TM$ be a Killing vector
field. The energy-flow of the gravitational potentials $\left(g_{\alpha}\right)$
along $\xi$ is the $1$--form $\mathcal{W}_{\xi}\in%TCIMACRO{\tbigwedge^{1}}%
%BeginExpansion
{\textstyle \bigwedge^{1}}%EndExpansion
M$ given by
\[
\mathcal{W}_{\xi}=\frac{1}{2}\star\left(dg_{\beta}\wedge i_{\xi}\star dg^{\beta}-i_{\xi}dg_{\beta}\wedge\star dg^{\beta}\right)\text{,}
\]
while $\delta\mathcal{W}_{\xi}=\mathcal{F}_{\xi}$ is the gravitational
force along $\xi$. \end{definition}

\begin{remark} Suppose that the dual $E_{\alpha}$ of a gravitational
potential $g_{\alpha}$ turns out to be a Killing vector field. So
the energy-flow $\mathcal{W}_{\xi}$ of the gravitational potentials
along $\xi=E_{\alpha}$ is just the first term of the gravitational
energy-momentum currents of \emph{Eq.(\ref{GEM})}. In fact, if the
Lorenz gauge $\delta g_{\beta}=0$ \emph{(}$0\leq\beta\leq3$\emph{)}
is assumed, $\mathcal{W}_{\xi}=\mathcal{T}_{\alpha}$. \end{remark}

\begin{lemma} Let $\left(g_{\alpha}\right)$ be gravitational potentials
obeying \emph{Eq.(\ref{EINSTEIN EQ})}, $\xi\in\sec TM$ a Killing
vector field and $\mathcal{W}_{\xi}$ the energy-flow of $\left(g_{\alpha}\right)$
along $\xi$. So
\[
\mathcal{F}_{\xi}=\delta\mathcal{W}_{\xi}=\left\langle i_{\xi}dg_{\alpha}|\mathcal{T}_{\alpha}+\mathcal{J}_{\alpha}\right\rangle \text{,}
\]
where $\left(\mathcal{T}_{\alpha}\right)$ and $\left(\mathcal{J}_{\alpha}\right)$
are the gravitational and matter energy-momentum currents, respectively.
\end{lemma}

Indeed, denoting by $\mathrm{L}_{\xi}$ the Lie derivative along $\xi$
and applying the Cartan's formula $\mathrm{L}_{\xi}=di_{\xi}+i_{\xi}d$,
\begin{align*}
d\star\mathcal{W}_{\xi} & =\frac{1}{2}di_{\mathrm{\xi}}\left(dg_{\alpha}\wedge\star dg^{\alpha}\right)-dg_{\alpha}\wedge di_{\mathrm{\xi}}\star dg^{\alpha}\\
 & =\frac{1}{2}\mathrm{L}_{\xi}\left(dg_{\alpha}\wedge\star dg^{\alpha}\right)-dg_{\alpha}\wedge di_{\mathrm{\xi}}\star dg^{\alpha}\\
 & =dg_{\alpha}\wedge\mathrm{L}_{\xi}\star dg^{\alpha}-dg_{\alpha}\wedge di_{\mathrm{\xi}}\star dg^{\alpha}\\
 & =dg_{\alpha}\wedge i_{\xi}d\star dg^{\alpha}\\
 & =-dg_{\alpha}\wedge i_{\xi}\star\delta dg^{\alpha}\text{.}
\end{align*}
From Einstein equation,
\begin{align*}
d\star\mathcal{W}_{\xi} & =-dg_{\alpha}\wedge i_{\xi}\star\left(\mathcal{T}_{\alpha}+\mathcal{J}_{\alpha}\right)\\
 & =i_{\xi}dg_{\alpha}\wedge\star\left(\mathcal{T}_{\alpha}+\mathcal{J}_{\alpha}\right)\text{,}
\end{align*}
so that
\[
\delta\mathcal{W}_{\xi}=-\star\left(i_{\xi}dg_{\alpha}\wedge\star\left(\mathcal{T}_{\alpha}+\mathcal{J}_{\alpha}\right)\right)=\left\langle i_{\xi}dg_{\alpha}|\mathcal{T}^{\alpha}+\mathcal{J}^{\alpha}\right\rangle \text{. }\blacksquare
\]

\begin{remark} \label{LORENTZ}The above force law for the gravitational
interaction with matter currents is analogous to the Lorentz force
of electrodynamics \cite{Thirring}. Indeed, let $\mathbf{F}\in%TCIMACRO{\tbigwedge^{2}}%
%BeginExpansion
{\textstyle \bigwedge^{2}}%EndExpansion
M$ be an electromagnetic field interacting with the current $\mathcal{J\in}%TCIMACRO{\tbigwedge^{1}}%
%BeginExpansion
{\textstyle \bigwedge^{1}}%EndExpansion
M$ according to the Maxwell inhomogeneous equation $\delta\mathbf{F}=\mathcal{J}$.
Following the same steps of the proof of the above Lemma, one can
show that $\mathcal{J}$ obeys the force law
\[
\delta\mathrm{T}_{\xi}=\left\langle i_{\xi}\mathbf{F}|\mathcal{J}\right\rangle \text{.}
\]
Here, $\mathrm{T}_{\xi}$ is the electromagnetic energy-momentum current
along $\xi$, given by
\[
\mathrm{T}_{\xi}=-\frac{1}{2}\star\left(\mathbf{F}\wedge i_{\xi}\star\mathbf{F}-i_{\xi}\mathbf{F}\wedge\star\mathbf{F}\right)\text{.}
\]
Compare $\mathrm{T}_{\xi}$ with the discussion of Remark \ref{ELECTROMAG}.
\end{remark}

\subsection{Gravitation as Non-Metricity\label{NM}}

On what follows, we rewrite our field equations (Eq.(\ref{EINSTEIN EQ}))
entirely in terms of the components $\mathbf{Q}_{\alpha\beta\gamma}$
of the non-metricity $2$--forms of the NM connection $\mathfrak{D}$
to which our gravitational potentials $\left(g_{\alpha}\right)$ are
adapted, together with the matter energy-momentum currents. This will
show that the gravitational field can be interpreted as the manifestation
of the non-metricity of $\mathfrak{D}$.

\begin{lemma} \label{NML}Let $\left(g_{\alpha}\right)\in%TCIMACRO{\tbigwedge}%
%BeginExpansion
{\textstyle \bigwedge}%EndExpansion
\mathfrak{D}$ and $\mathbf{Q}_{\alpha\beta\gamma}$ be the components of the non-metricity
$2$--forms $\left(\mathbf{Q}_{\gamma}\right)$ of $\mathfrak{D}$
relative to $\left(g_{\alpha}\right)$. So,
\begin{equation}
i_{\nu}\left(\delta dg_{\mu}\right)=\mathbf{Q}_{\mu\lbrack\alpha\nu];}^{~~~~~~\alpha}-\mathbf{Q}_{\mu\alpha\beta}\mathbf{Q}_{\nu}^{~[\alpha\beta]}-\mathbf{Q}_{\mu\lbrack\alpha\nu]}\mathbf{Q}_{~\beta}^{\alpha~\beta}\text{.}\label{I1}
\end{equation}
Also, let $\left(\mathcal{T}_{\alpha}\right)\in%TCIMACRO{\tbigwedge^{1}}%
%BeginExpansion
{\textstyle \bigwedge^{1}}%EndExpansion
M$ be the gravitational energy-momentum currents \emph{(Eq.(\ref{GEM}))
}of the gravitational potentials $\left(g_{\alpha}\right)$. Hence,
\begin{align}
i_{\nu}\mathcal{T}_{\mu} & =\mathbf{Q}_{\alpha\lbrack\mu\beta]}\mathbf{Q}^{\alpha\lbrack\beta\gamma]}\eta_{\gamma\nu}-\mathbf{Q}_{\alpha\nu~;\mu}^{~~\alpha}\nonumber \\
 & +\left[\mathbf{Q}_{\alpha\beta}^{~~\alpha;\beta}+\frac{1}{2}\left(\mathbf{Q}_{\alpha\beta\gamma}\mathbf{Q}^{\alpha\lbrack\beta\gamma]}+\mathbf{Q}_{\alpha\beta}^{~~\alpha}\mathbf{Q}_{\gamma}^{~\beta\gamma}\right)\right]\eta_{\mu\nu}\text{.}\label{I2}
\end{align}

\end{lemma}

First we prove Eq.(\ref{I1}). Let $(\theta_{~\beta}^{\alpha})$ be
the Levi-Civita connection $1$--forms of the pseudo-Riemannian space
$\left(M,\mathbf{g}\right)$, where $\mathbf{g}=\eta_{\alpha\beta}g^{\alpha}\otimes g^{\beta}$.
By Corollary \ref{Non-metricity C.}, 
\begin{equation}
dg_{\mu}=-\theta_{\mu\alpha}\wedge g^{\alpha}=-\left(\mathbf{Q}_{\mu\alpha\beta}g^{\beta}\right)\wedge g^{\alpha}=\mathbf{Q}_{\mu\alpha\beta}g^{\alpha}\wedge g^{\beta}\text{.}\label{J0}
\end{equation}
So,
\begin{align}
\star d\star\left(dg_{\mu}\right) & =\star d\left(\mathbf{Q}_{\mu\alpha\beta}\star\left(g^{\alpha}\wedge g^{\beta}\right)\right)\nonumber \\
 & =\star\left(d\mathbf{Q}_{\mu\alpha\beta}\wedge\star\left(g^{\alpha}\wedge g^{\beta}\right)\right)+\mathbf{Q}_{\mu\alpha\beta}\star d\star\left(g^{\alpha}\wedge g^{\beta}\right)\nonumber \\
 & =\mathbf{Q}_{\mu\alpha\beta;\delta}\star\left(g^{\delta}\wedge\star\left(g^{\alpha}\wedge g^{\beta}\right)\right)+\mathbf{Q}_{\mu\alpha\beta}\star d\star\left(g^{\alpha}\wedge g^{\beta}\right)\text{.}\label{J1}
\end{align}
On one hand,
\[
\star\left(g^{\delta}\wedge\star\left(g^{\alpha}\wedge g^{\beta}\right)\right)=-i^{\delta}\left(g^{\alpha}\wedge g^{\beta}\right)=\eta^{\beta\delta}g^{\alpha}-\eta^{\alpha\delta}g^{\beta}\text{,}
\]
and therefore
\begin{align}
\mathbf{Q}_{\mu\alpha\beta;\delta}\star\left(g^{\delta}\wedge\star\left(g^{\alpha}\wedge g^{\beta}\right)\right) & =\mathbf{Q}_{\mu\alpha\beta;\delta}\eta^{\beta\delta}g^{\alpha}-\mathbf{Q}_{\mu\alpha\beta;\delta}\eta^{\alpha\delta}g^{\beta}\nonumber \\
 & =\mathbf{Q}_{\mu\alpha\beta;}^{~~~~\beta}g^{\alpha}-\mathbf{Q}_{\mu\alpha\beta;}^{~~~~\alpha}g^{\beta}\nonumber \\
 & =(\mathbf{Q}_{\mu\alpha\beta;}^{~~~~\beta}-\mathbf{Q}_{\mu\beta\alpha;}^{~~~~\beta})g^{\alpha}\nonumber \\
 & =\mathbf{Q}_{\mu\left[\alpha\beta\right];}^{~~~~~~\beta}g^{\alpha}\text{.}\label{J2}
\end{align}
Now, on the other hand,
\begin{align*}
d\star\left(g^{\alpha}\wedge g^{\beta}\right) & =-\theta_{~\gamma}^{\alpha}\wedge\star\left(g^{\gamma}\wedge g^{\beta}\right)-\theta_{~\gamma}^{\beta}\wedge\star\left(g^{\alpha}\wedge g^{\gamma}\right)\\
 & =-\mathbf{Q}_{~\gamma\delta}^{\alpha}g^{\delta}\wedge\star\left(g^{\gamma}\wedge g^{\beta}\right)-\mathbf{Q}_{~\gamma\delta}^{\beta}g^{\delta}\wedge\star\left(g^{\alpha}\wedge g^{\gamma}\right)\text{,}
\end{align*}
using Corollary \ref{Non-metricity C.} again. Thus
\[
\star d\star\left(g^{\alpha}\wedge g^{\beta}\right)=-\mathbf{Q}_{~\gamma\delta}^{\alpha}\star\left(\star\left(g^{\gamma}\wedge g^{\beta}\right)\wedge g^{\delta}\right)-\mathbf{Q}_{~\gamma\delta}^{\beta}\star\left(\star\left(g^{\alpha}\wedge g^{\gamma}\right)\wedge g^{\delta}\right)\text{.}
\]
But since
\begin{align*}
\star\left(\star\left(g^{\gamma}\wedge g^{\beta}\right)\wedge g^{\delta}\right) & =-i^{\delta}\left(g^{\gamma}\wedge g^{\beta}\right)=\eta^{\beta\delta}g^{\gamma}-\eta^{\gamma\delta}g^{\beta}\text{,}\\
\star\left(\star\left(g^{\alpha}\wedge g^{\gamma}\right)\wedge g^{\delta}\right) & =-i^{\delta}\left(g^{\alpha}\wedge g^{\gamma}\right)=\eta^{\gamma\delta}g^{\alpha}-\eta^{\alpha\delta}g^{\gamma}\text{,}
\end{align*}
we obtain:
\begin{align*}
\star d\star\left(g^{\alpha}\wedge g^{\beta}\right) & =-\mathbf{Q}_{~\gamma\delta}^{\alpha}\left(\eta^{\beta\delta}g^{\gamma}-\eta^{\gamma\delta}g^{\beta}\right)-\mathbf{Q}_{~\gamma\delta}^{\beta}\left(\eta^{\gamma\delta}g^{\alpha}-\eta^{\alpha\delta}g^{\gamma}\right)\\
 & =\mathbf{Q}_{~\gamma}^{\alpha~\gamma}g^{\beta}-\mathbf{Q}_{~\gamma}^{\beta~\gamma}g^{\alpha}+(\mathbf{Q}_{~\gamma}^{\beta~\alpha}-\mathbf{Q}_{~\gamma}^{\alpha~\beta})g^{\gamma}\\
 & =\mathbf{Q}_{~\gamma}^{\alpha~\gamma}g^{\beta}-\mathbf{Q}_{~\gamma}^{\beta~\gamma}g^{\alpha}+(\mathbf{Q}_{\gamma}^{~\alpha\beta}-\mathbf{Q}_{\gamma}^{~\beta\alpha})g^{\gamma}\\
 & =\mathbf{Q}_{~\gamma}^{\alpha~\gamma}g^{\beta}-\mathbf{Q}_{~\gamma}^{\beta~\gamma}g^{\alpha}+\mathbf{Q}_{\gamma}^{~\left[\alpha\beta\right]}g^{\gamma}\text{.}
\end{align*}
So, multiplying by $\mathbf{Q}_{\mu\alpha\beta}$ yields
\begin{align}
\mathbf{Q}_{\mu\alpha\beta}\star d\star\left(g^{\alpha}\wedge g^{\beta}\right) & =\mathbf{Q}_{\mu\alpha\beta}\mathbf{Q}_{\gamma}^{~\left[\alpha\beta\right]}g^{\gamma}+\mathbf{Q}_{\mu\alpha\beta}\mathbf{Q}_{~\gamma}^{\alpha~\gamma}g^{\beta}-\mathbf{Q}_{\mu\alpha\beta}\mathbf{Q}_{~\gamma}^{\beta~\gamma}g^{\alpha}\nonumber \\
 & =\mathbf{Q}_{\mu\alpha\beta}\mathbf{Q}_{\gamma}^{~\left[\alpha\beta\right]}g^{\gamma}+(\mathbf{Q}_{\mu\alpha\beta}-\mathbf{Q}_{\mu\beta\alpha})\mathbf{Q}_{~\gamma}^{\alpha~\gamma}g^{\beta}\nonumber \\
 & =\mathbf{Q}_{\mu\alpha\beta}\mathbf{Q}_{\gamma}^{~\left[\alpha\beta\right]}g^{\gamma}+\mathbf{Q}_{\mu\left[\alpha\beta\right]}\mathbf{Q}_{~\gamma}^{\alpha~\gamma}g^{b}\text{.}\label{J3}
\end{align}
Hence, from Eqs.(\ref{J1}), (\ref{J2}) and (\ref{J3}), we deduce:
\[
\star d\star\left(dg_{\mu}\right)=\mathbf{Q}_{\mu\left[\alpha\beta\right];}^{~~~~~~\beta}g^{\alpha}+\mathbf{Q}_{\mu\alpha\beta}\mathbf{Q}_{\gamma}^{~\left[\alpha\beta\right]}g^{\gamma}+\mathbf{Q}_{\mu\left[\alpha\beta\right]}\mathbf{Q}_{~\gamma}^{\alpha~\gamma}g^{\beta}\text{.}
\]
By contracting with $i_{\nu}$,
\[
i_{\nu}\left(\star d\star\left(dg_{\mu}\right)\right)=\mathbf{Q}_{\mu\left[\nu\beta\right];}^{~~~~~~\beta}+\mathbf{Q}_{\mu\alpha\beta}\mathbf{Q}_{\nu}^{~\left[\alpha\beta\right]}+\mathbf{Q}_{\mu\left[\alpha\nu\right]}\mathbf{Q}_{~\gamma}^{\alpha~\gamma}\text{.}
\]
Finally, since $i_{\nu}\left(\delta dg_{\mu}\right)=-i_{\nu}\left(\star d\star\left(dg_{\mu}\right)\right)$,
we obtain
\[
i_{\nu}\left(\delta dg_{\mu}\right)=\mathbf{Q}_{\mu\left[\alpha\nu\right];}^{~~~~~~\alpha}-\mathbf{Q}_{\mu\alpha\beta}\mathbf{Q}_{\nu}^{~\left[\alpha\beta\right]}-\mathbf{Q}_{\mu\left[\alpha\nu\right]}\mathbf{Q}_{~\beta}^{\alpha~\beta}\text{.}
\]

Now we derive Eq.(\ref{I2}). Using that
\[
dg_{\alpha}\wedge i_{\mu}\star dg^{\alpha}=\left\langle dg_{\alpha}|dg^{\alpha}\right\rangle \star g_{\mu}-i_{\mu}dg_{\alpha}\wedge\star dg^{\alpha}\text{,}
\]
the gravitational energy-momentum currents (Eq.(\ref{GEM})) can be
written as
\begin{align*}
\mathcal{T}_{\mu} & =\frac{1}{2}\left\langle dg_{\alpha}|dg^{\alpha}\right\rangle g_{\mu}+i_{i_{\mu}dg_{\alpha}}dg^{\alpha}\\
 & +\frac{1}{2}\delta g_{\alpha}\wedge\delta g^{\alpha}\wedge g_{\mu}+i_{\alpha}d\delta g^{\alpha}\wedge g_{\mu}-i_{\mu}d\delta g^{\alpha}\wedge g_{\alpha}\text{.}
\end{align*}
Contraction with $i_{\nu}$ yields
\begin{align}
i_{\nu}\mathcal{T}_{\mu} & =\left(\frac{1}{2}\left\langle dg_{\alpha}|dg^{\alpha}\right\rangle +\frac{1}{2}\delta g_{\alpha}\wedge\delta g^{\alpha}+i_{\alpha}d\delta g^{\alpha}\right)\eta_{\mu\nu}\nonumber \\
 & +i_{\nu}\left(i_{i_{\mu}dg_{\alpha}}dg^{\alpha}\right)-i_{\mu}d\delta g_{\nu}\text{.}\label{J4}
\end{align}

First, use Eq.(\ref{J0}) to obtain
\[
\left\langle dg_{\alpha}|dg^{\alpha}\right\rangle =\mathbf{Q}_{\alpha\beta\gamma}\mathbf{Q}_{~\delta\epsilon}^{\alpha}\left\langle g^{\beta}\wedge g^{\gamma}|g^{\delta}\wedge g^{\epsilon}\right\rangle \text{.}
\]
But
\begin{align*}
\left\langle g^{\beta}\wedge g^{\gamma}|g^{\delta}\wedge g^{\epsilon}\right\rangle  & =i^{\gamma}i^{\beta}\left(g^{\delta}\wedge g^{\epsilon}\right)\\
 & =i^{\gamma}\left(\eta^{\beta\delta}g^{\epsilon}-\eta^{\beta\epsilon}g^{\delta}\right)=\eta^{\beta\delta}\eta^{\gamma\epsilon}-\eta^{\beta\epsilon}\eta^{\gamma\delta}
\end{align*}
implies
\begin{align}
\left\langle dg_{\alpha}|dg^{\alpha}\right\rangle  & =\mathbf{Q}_{\alpha\beta\gamma}\mathbf{Q}_{~\delta\epsilon}^{\alpha}\eta^{\beta\delta}\eta^{\gamma\epsilon}-\mathbf{Q}_{\alpha\beta\gamma}\mathbf{Q}_{~\delta\epsilon}^{\alpha}\eta^{\beta\epsilon}\eta^{\gamma\delta}\nonumber \\
 & =\mathbf{Q}_{\alpha\beta\gamma}\left(\mathbf{Q}^{\alpha\beta\gamma}-\mathbf{Q}^{\alpha\gamma\beta}\right)=\mathbf{Q}_{\alpha\beta\gamma}\mathbf{Q}^{\alpha\left[\beta\gamma\right]}\text{.}\label{J5}
\end{align}

Second, from Lemma \ref{Codiff.} and Corollary \ref{Non-metricity C.},
\[
\delta g_{\alpha}=i^{\beta}\theta_{\beta\alpha}=i^{\beta}\left(\mathbf{Q}_{\beta\alpha\gamma}g^{\gamma}\right)=\mathbf{Q}_{\beta\alpha}^{~~\beta}\text{,}
\]
so that
\begin{equation}
\delta g_{\alpha}\wedge\delta g^{\alpha}=\mathbf{Q}_{\beta\alpha}^{~~\beta}\mathbf{Q}_{\gamma}^{~\alpha\gamma}\text{.}\label{J6}
\end{equation}

Third,
\begin{equation}
i^{\alpha}d\delta g_{\alpha}=i^{\alpha}(\mathbf{Q}_{\beta\alpha~;\gamma}^{~~\beta}g^{\gamma})=\mathbf{Q}_{\alpha\beta~;}^{~~\alpha~\beta}\label{J7}
\end{equation}
and
\begin{equation}
i_{\mu}d\delta g_{\nu}=i_{\mu}(\mathbf{Q}_{\beta\nu~;\gamma}^{~~\beta}g^{\gamma})=\mathbf{Q}_{\alpha\nu~;\mu}^{~~\alpha}\text{.}\label{J8}
\end{equation}

Lastly, from Eq.(\ref{J0}),
\begin{align*}
i_{\mu}dg_{\alpha} & =i_{\mu}(\mathbf{Q}_{\alpha\beta\gamma}g^{\beta}\wedge g^{\gamma})=\mathbf{Q}_{\alpha\beta\gamma}(\delta_{\mu}^{\beta}g^{\gamma}-\delta_{\mu}^{\gamma}g^{\beta})\\
 & =(\mathbf{Q}_{\alpha\mu\beta}-\mathbf{Q}_{\alpha\beta\mu})g^{\beta}=\mathbf{Q}_{\alpha\left[\mu\beta\right]}g^{\beta}\text{,}
\end{align*}
which imply
\begin{align*}
i_{i_{\mu}dg_{\alpha}}dg^{\alpha} & =\mathbf{Q}_{\alpha\left[\mu\beta\right]}\mathbf{Q}_{~\gamma\delta}^{\alpha}i^{\beta}(g^{\gamma}\wedge g^{\delta})\\
 & =\mathbf{Q}_{\alpha\left[\mu\beta\right]}\mathbf{Q}_{~\gamma\delta}^{\alpha}\left(\eta^{\beta\gamma}g^{\delta}-\eta^{\beta\delta}g^{\gamma}\right)\\
 & =\mathbf{Q}_{\alpha\left[\mu\beta\right]}\mathbf{Q}^{\alpha\left[\beta\gamma\right]}g_{\gamma}\text{.}
\end{align*}
Hence:
\begin{equation}
i_{\nu}\left(i_{i_{\mu}dg_{\alpha}}dg^{\alpha}\right)=\mathbf{Q}_{\alpha\left[\mu\beta\right]}\mathbf{Q}^{\alpha\left[\beta\gamma\right]}\eta_{\gamma\nu}\text{.}\label{J9}
\end{equation}

Finally, substituting Eqs.(\ref{J5}) to (\ref{J9}) in Eq.(\ref{J4}),
we conclude:
\begin{align}
i_{\nu}\mathcal{T}_{\mu} & =\mathbf{Q}_{\alpha\left[\mu\beta\right]}\mathbf{Q}^{\alpha\left[\beta\gamma\right]}\eta_{\gamma\nu}-\mathbf{Q}_{\alpha\nu~;\mu}^{~~\alpha}\\
 & +\left[\mathbf{Q}_{\alpha\beta~;}^{~~\alpha~\beta}+\frac{1}{2}\left(\mathbf{Q}_{\alpha\beta\gamma}\mathbf{Q}^{\alpha\left[\beta\gamma\right]}+\mathbf{Q}_{\beta\alpha}^{~~\beta}\mathbf{Q}_{\gamma}^{~\alpha\gamma}\right)\right]\eta_{\mu\nu}\text{. }\blacksquare
\end{align}

\begin{proposition} The Einstein equations \emph{(Eq.(\ref{EINSTEIN EQ}))}
for the gravitational potentials $\left(g_{\alpha}\right)\in%TCIMACRO{\tbigwedge}%
%BeginExpansion
{\textstyle \bigwedge}%EndExpansion
\mathfrak{D}$ coupled to the matter energy-momentum currents $\left(\mathcal{J}_{\alpha}\right)$
are equivalent to the following field equations for the components
$\mathbf{Q}_{\alpha\beta\gamma}$ of the non-metricity $2$--forms
$\left(\mathbf{Q}_{\gamma}\right)$ of $\mathfrak{D}$ relative to
$\left(g_{\alpha}\right)$,
\begin{align}
i_{\nu}\mathcal{J}_{\mu} & =\mathbf{Q}_{\mu\lbrack\alpha\nu];}^{~~~~~~\alpha}+\mathbf{Q}_{\alpha\nu~;\mu}^{~~\alpha}-\left[\mathbf{Q}_{\alpha\beta}^{~~\alpha;\beta}+\frac{1}{2}\left(\mathbf{Q}_{\alpha\beta\gamma}\mathbf{Q}^{\alpha\lbrack\beta\gamma]}+\mathbf{Q}_{\alpha\beta}^{~~\alpha}\mathbf{Q}_{\gamma}^{~\beta\gamma}\right)\right]\eta_{\mu\nu}\nonumber \\
 & -\mathbf{Q}_{\alpha\lbrack\mu\beta]}\mathbf{Q}^{\alpha\lbrack\beta\gamma]}\eta_{\gamma\nu}-\mathbf{Q}_{\mu\alpha\beta}\mathbf{Q}_{\nu}^{~[\alpha\beta]}-\mathbf{Q}_{\mu\lbrack\alpha\nu]}\mathbf{Q}_{~\beta}^{\alpha~\beta}\text{.}\label{NM FIELD EQ}
\end{align}

\end{proposition}

The result follows from a direct substitution of Eqs.(\ref{I1}) and
(\ref{I2}) in the Einstein equations (Eq.(\ref{EINSTEIN EQ})). $\blacksquare$

\begin{remark} The linearized gravitational field equations in terms
of the components $\mathbf{Q}_{\alpha\beta\gamma}$ of the non-metricity
$2$--forms are
\[
i_{\nu}\mathcal{J}_{\mu}=\mathbf{Q}_{\mu\lbrack\alpha\nu];}^{~~~~~~\alpha}+\mathbf{Q}_{\alpha\nu~;\mu}^{~~\alpha}-\mathbf{Q}_{\alpha\beta}^{~~\alpha;\beta}\text{,}
\]
as seen easily from \emph{Eq.(\ref{NM FIELD EQ})}. They constitute
a coupled system of first order partial differential equations. \end{remark}

\begin{example} {[}Schwarzschild solution{]}Let $m>0$, $M=\mathbb{R\times}\left]2m,\infty\right[\times S^{2}$
and $\left(t,r,\theta,\varphi\right)$ be the natural coordinates
of $M$. Also, let
\[
\mathbf{g}=-\left(1-\frac{2m}{r}\right)dt\otimes dt+\frac{1}{1-2m/r}dr\otimes dr+r^{2}\omega\text{,}
\]
where $\omega$ is the pull-back of the Euclidean metric of $S^{2}$,
\[
\omega=d\theta\otimes d\theta+\sin^{2}\theta d\varphi\otimes d\varphi\text{.}
\]
Then $\left(M,\mathbf{g}\right)$ is called the Schwarzschild solution
with mass $m$, and $\left(t,r,\theta,\varphi\right)$ are known as
Schwarzschild coordinates \cite{Mol}. The gravitational potentials
of $\left(M,\mathbf{g}\right)$ can be described by the coframe field
$\left(g_{\alpha}\right)$ such that
\[
g^{0}=\sqrt{1-\frac{2m}{r}}dt\text{, \ \ }g^{1}=\frac{1}{\sqrt{1-2m/r}}dr\text{, \ \ }g^{2}=rd\theta\text{, \ \ }g^{3}=r\sin\theta d\varphi\text{.}
\]
A NM connection can be defined for which $\left(g_{\alpha}\right)$
is an adapted coframe field, once we choose its non-metricity $1$--forms
$(\mathcal{A}_{~\beta}^{\alpha})$ as
\[
\mathcal{A}_{~1}^{0}=\frac{1}{\sqrt{1-2m/r}}\frac{m}{r^{2}}g^{0}\text{, \ \ }\mathcal{A}_{~1}^{2}=\frac{1}{r}\sqrt{1-\frac{2m}{r}}g^{2}\text{,}
\]
\[
\mathcal{A}_{~1}^{3}=\frac{1}{r}\sqrt{1-\frac{2m}{r}}g^{3}\text{, \ \ }\mathcal{A}_{~2}^{3}=\frac{1}{r\tan\theta}g^{3}\text{,}
\]
and with diagonal elements
\[
\mathcal{A}_{~0}^{0}=F_{0}g^{0}\text{, ..., }\mathcal{A}_{~3}^{3}=F_{3}g^{3}\text{,}
\]
where $\left(F_{\alpha}\right)$ are any differentiable functions
$M\longrightarrow\mathbb{R}$. It is easily proven that 
\[
dg_{\alpha}=-\mathcal{A}_{\alpha\beta}\wedge g^{\alpha},
\]
so that our NM is well-defined. \end{example}

\begin{remark} A discussion of the Schwarzschild solution in terms
of non-metricity is also presented by Notte-Cuello, da Rocha and Rodrigues
in \cite{NRR}. However, instead of using a non-metricity gauge or
NM connections, these authors considered the situation in which the
gravitational field is derived from the non-metricity of the Levi-Civita
connection compatible with a Minkowski metric, defined over the Schwarzschild
spacetime. Therefore, they work requires a bimetric theory of gravitation,
while our theory only requires a cobase satisfying the non-metricity
gauge. \end{remark}

\section{Discussion\label{Conclusion}}

In trying to better understand the gauge nature of gravitation, Thirring
and Wallner \cite{Thirring} \cite{W} \cite{TW} were led to a gravitational
Lagrangian density involving only the cobase that represents the gravitational
potentials (recall Definition \ref{WL} and Remark \ref{TW}). Their
approach, as the reader may be convinced by studying our Appendix
\ref{EHL}, is not restricted to any geometrical interpretation of
the gravitational field. The latter is then realized as a \textit{legitimate}
field living in the spacetime manifold, in a sense similar to that
with which Faraday, Maxwell and Lorentz attributed to the electromagnetism.

Now, concerned with the existence of conservation laws in GR and guided
by Thirring and Wallner writings, Rodrigues and his collaborators
\cite{Rodrigues} \cite{RF} \cite{RQ} \cite{R} proposed to rewrite
the Einstein equations coupled to the matter currents $\left(\mathcal{J}_{\alpha}\right)$
as
\[
\delta dg_{\alpha}=\mathcal{T}_{\alpha}+\mathcal{J}_{\alpha}\text{,}
\]
where the $1$--forms $\left(\mathcal{T}_{\alpha}\right)$ are identified
with the gravitational energy-momentum currents. This parallels the
equations of gravitation with the Maxwell inhomogeneous equation of
electrodynamics, and realizes the physical idea that the gravitational
energy-momentum currents are itself a source for gravitational fields.

On the other hand, many authors have been concerned with new geometrical
interpretations of GR, principally in teleparallel spaces. In particular,
Nester \cite{Nester}, Adak and his collaborators \cite{Adak1}--\cite{Adak4}
studied the geometrical formulation of GR based on a flat torsionless
connection, where the gravitational field is manifest in the non-metricity
of such a connection. This approach, which we have called the ``non-metricity
formulation of GR\textquotedblright , has been referred by the latter
authors (starting with Nester) as the ``Symmetric Teleparallel General
Relativity\textquotedblright \ (STGR). \label{Nester}

In the present paper, we have unified the works of Thirring, Wallner,
Rodrigues, Nester, Adak and many others. We begun with the gravitational
formalism of Thirring and Wallner, conceiving the gravitational potentials
as a cobase field $\left(g_{\alpha}\right)$ living in a parallelizable
spacetime manifold. Then, from our discussion of the NM connections,
we introduced the notion of the non-metricity gauge, for which the
gravitational potentials satisfy
\[
g_{\alpha}\wedge dg^{\alpha}=0\text{.}
\]
The geometrical interpretation of this gauge is that the cobase $\left(g_{\alpha}\right)$
is \textit{adapted} to a given NM connection $\mathfrak{D}$, which
means that the connection $1$--forms of $\mathfrak{D}$ coincides
with its non-metricity $1$--forms relative to $\left(g_{\alpha}\right)$.

Then the Wallner Lagrangian density, when restricted to the non-metricity
gauge, becomes
\[
\mathcal{L}|_{\wedge\mathfrak{D}}=\frac{1}{2}g_{\alpha}\wedge dg^{\beta}\wedge\star\left(g_{\beta}\wedge dg^{\alpha}\right)\text{.}
\]
By employing the variational principle and following Rodrigues remarks,
we obtained the gravitational field equations coupled to the matter
energy-momentum currents $\left(\mathcal{J}_{\alpha}\right)$ as $\delta dg_{\alpha}=\mathcal{T}_{\alpha}+\mathcal{J}_{\alpha}$.
However, from our concern with the non-metricity formulation of gravitation,
we found that the gravitational energy-momentum currents in the non-metricity
gauge assumes a particularly simple and physically appealing form,
namely,
\begin{align*}
\mathcal{T}_{\alpha} & =\frac{1}{2}\star\left(dg_{\beta}\wedge i_{\alpha}\star dg^{\beta}-i_{\alpha}dg_{\beta}\wedge\star dg^{\beta}\right)\\
 & +\frac{1}{2}\delta g_{\beta}\wedge\delta g^{\beta}\wedge g_{\alpha}+i_{\beta}d\delta g^{\beta}\wedge g_{\alpha}-i_{\alpha}d\delta g^{\beta}\wedge g_{\beta}\text{,}
\end{align*}
principally if compared with Rodrigues' original expression for $\left(\mathcal{T}_{\alpha}\right)$
\cite{R}.

From this, we could deduce that if the gravitational Lorenz gauge
is assumed, for which 
\[
\delta g_{\alpha}=0\text{, \ \ }0\leq\alpha\leq3\text{,}
\]
the gravitational field equations becomes a system of four coupled
Proca equations with variable mass,
\[
\square g_{\alpha}+\frac{1}{2}\left\langle dg_{\beta}|dg^{\beta}\right\rangle g_{\alpha}=i_{dg_{\beta}}\left(g_{\alpha}\wedge dg^{\beta}\right)+\mathcal{J}_{\alpha}\text{.}
\]
As we have indicated in Examples \ref{WAVES 1} and \ref{WAVES 2},
these equations may be of interest in the study of the propagation
of gravitational-electromagnetic waves, as illustrated in the solutions
of \cite{Bramson}, \cite{Vaidya} and \cite{Roy}\emph{.}

As another consequence of our study of the gravitational equations,
we proved a particularly simple force law for the matter currents
coupled to the gravitational field. Namely, that if we identify the
$1$--form 
\[
\mathcal{W}_{\xi}=\frac{1}{2}\star\left(dg_{\beta}\wedge i_{\xi}\star dg^{\beta}-i_{\xi}dg_{\beta}\wedge\star dg^{\beta}\right)
\]
with the gravitational energy-flow along the Killing vector field
$\xi\in\sec TM$, then
\[
\delta\mathcal{W}_{\xi}=\left\langle i_{\xi}dg_{\alpha}|\mathcal{T}^{\alpha}+\mathcal{J}^{\alpha}\right\rangle \text{.}
\]
It must be observed that by employing the same argument used in the
proof of Lemma \ref{NML}, one can express the above force law entirely
in terms of the components of the $2$--form of non-metricity together
with the matter currents $\left(\mathcal{J}_{\alpha}\right)$. In
this way, the coupling of matter with non-metricity in our formalism
can still be analyzed in more details.

Finally, as Nester, Adak and collaborators, we proved that a gravitational
theory equivalent to GR can be formulated so that the gravitational
field derives purely from the non-metricity of a flat torsionless
connection. Particularly, we showed that our gravitational field equations
coupled to the matter currents $\left(\mathcal{J}_{\alpha}\right)$
can be rewritten as
\begin{align*}
i_{\nu}\mathcal{J}_{\mu} & =\mathbf{Q}_{\mu\lbrack\alpha\nu];}^{~~~~~~\alpha}+\mathbf{Q}_{\alpha\nu~;\mu}^{~~\alpha}-\left[\mathbf{Q}_{\alpha\beta}^{~~\alpha;\beta}+\frac{1}{2}\left(\mathbf{Q}_{\alpha\beta\gamma}\mathbf{Q}^{\alpha\lbrack\beta\gamma]}+\mathbf{Q}_{\alpha\beta}^{~~\alpha}\mathbf{Q}_{\gamma}^{~\beta\gamma}\right)\right]\eta_{\mu\nu}\\
 & -\mathbf{Q}_{\alpha\lbrack\mu\beta]}\mathbf{Q}^{\alpha\lbrack\beta\gamma]}\eta_{\gamma\nu}-\mathbf{Q}_{\mu\alpha\beta}\mathbf{Q}_{\nu}^{~[\alpha\beta]}-\mathbf{Q}_{\mu\lbrack\alpha\nu]}\mathbf{Q}_{~\beta}^{\alpha~\beta}\text{,}
\end{align*}
where $\mathbf{Q}_{\alpha\beta\gamma}$ are the components of the
non-metricity $2$--form of the NM connection $\mathfrak{D}$ to which
our gravitational potentials are adapted. From this, we see that the
linearized field equations in terms of non-metricity are
\[
i_{\nu}\mathcal{J}_{\mu}=\mathbf{Q}_{\mu\lbrack\alpha\nu];}^{~~~~~~\alpha}+\mathbf{Q}_{\alpha\nu~;\mu}^{~~\alpha}-\mathbf{Q}_{\alpha\beta}^{~~\alpha;\beta}\text{,}
\]
which constitute a coupled system of first order partial differential
equations.

\bigskip{}

Now we close by commenting the following remark by Nester \cite{Nester}.
\begin{quotation}
``Of course the STGR\footnote{``Symmetric Teleparallel General Relativity\textquotedblright . See
p. \pageref{Nester}.} formulation has some liabilities. It must be emphasized that in this
geometry it is no longer possible to simply commute derivatives and
the raising or lowering of indices via the metric as we are so accustomed
to do in the standard Riemannian approach. Hence tensorial equations
will appear differently depending on how the indices are arranged.
(...) Another obvious limitation of the STGR formulation is that it
(almost) requires a global coordinate system\textquotedblright . 
\end{quotation}
The above criticisms are irrelevant to our non-metricity formulation
of GR, as we have adopted the calculus of differential forms instead
of the classical tensorial calculus. We assumed that our spacetime
manifold is parallelizable (something which can be justified physically
from the existence of spinorial fields) and that the gravitational
potentials are represented by a cobase field $\left(g_{\alpha}\right)$.
The only components which we have utilized above are the components
$\mathbf{Q}_{\alpha\beta\gamma}$ relative to $\left(g_{\alpha}\right)$
of the $2$--form of non-metricity $\mathbf{Q}_{\gamma}$ in Eq.(\ref{NM FIELD EQ}),
for which
\[
\mathbf{Q}_{\gamma}=\frac{1}{2}\mathbf{Q}_{\alpha\beta\gamma}g^{\alpha}\wedge g^{\beta}\text{.}
\]
As $\left(g_{\alpha}\right)$ exists globally and our ``raising and
lowering\textquotedblright \ of indices derives from the metric $\mathbf{g}=\eta_{\alpha\beta}g^{\alpha}\otimes g^{\beta}$,
we can raise and lower indices in Eq.(\ref{NM FIELD EQ}) with $\mathbf{g}$
as usual, and no global coordinate system is required.

\bigskip{}

\textbf{Acknowledgment.} The author is grateful to Waldyr Rodrigues,
for the important discussions during the development of this work,
to Zbigniew Oziewicz, for having read and commented on the manuscript,
and to Yen Chin and Muzaffer Adak, for having called my attention
to the literature of the STGR.

\appendix
%dummy comment inserted by tex2lyx to ensure that this paragraph is not empty

\section{Einstein-Hilbert Lagrangian\label{EHL}}

In this Appendix, we show how the usual interpretation of GR in terms
of the curvature of the pseudo-Riemmanian space $(M,\mathbf{g,\nabla})$,
where $\mathbf{g}=\eta_{\alpha\beta}g^{\alpha}\otimes g^{\beta}$
is induced by the gravitational potentials $\left(g_{\alpha}\right)$,
arises from the Wallner Lagrangian. That is, we shall geometrize the
theory in a Lorentzian space by giving a privilege to its Levi-Civita
connection.

\bigskip{}

First, we show that the WL can be decomposed in three terms, one of
which is of Yang-Mills type. Such decomposition is due to Rodrigues
and de Souza \cite{RQ}.

\begin{lemma} \label{RdS}The WL can be written as
\[
\mathcal{L}=\frac{1}{2}dg_{\alpha}\wedge\star dg^{\alpha}-\frac{1}{2}\delta g_{\alpha}\wedge\star\delta g^{\alpha}-\frac{1}{4}g_{\alpha}\wedge dg^{\alpha}\wedge\star\left(g_{\beta}\wedge dg^{\beta}\right)\text{.}
\]

\end{lemma}

In fact, using Lemma \ref{Trick},
\begin{align*}
\delta g_{\alpha}\wedge\star\delta g^{\alpha} & =i_{\alpha}dg^{\alpha}\wedge\star i_{\beta}dg^{\beta}=-dg^{\alpha}\wedge i_{\alpha}\star i_{\beta}dg^{\beta}\\
 & =-dg^{\alpha}\wedge\star\left(i_{\beta}dg^{\beta}\wedge g_{\alpha}\right)\\
 & =-dg^{\alpha}\wedge\star i_{\beta}\left(dg^{\beta}\wedge g_{\alpha}\right)+dg^{\alpha}\wedge\star dg_{\alpha}\\
 & =-i_{\beta}\left(dg^{\beta}\wedge g_{\alpha}\right)\wedge\star dg^{\alpha}+dg^{\alpha}\wedge\star dg_{\alpha}\\
 & =-g_{\alpha}\wedge dg^{\beta}\wedge\star\left(g_{\beta}\wedge dg^{\alpha}\right)+dg^{\alpha}\wedge\star dg_{\alpha}\text{. }\blacksquare
\end{align*}

\bigskip{}

Now we geometrize the gravitational theory.

\begin{lemma} Let $M$ be a four--dimensional parallelizable manifold
and $\left(g_{\alpha}\right)$ a tetrad on $M$ such that $\mathbf{g}=\eta_{\alpha\beta}g^{\alpha}\otimes g^{\beta}$
is a Lorentzian metric. Let $(\mathcal{R}_{~\beta}^{\alpha})$ be
the curvature $2$--forms of the Levi-Civita connection of $\left(M,\mathbf{g}\right)$,
$\left(\mathcal{R}_{\alpha}\right)$ the Ricci $1$--forms and $\mathcal{R}=i_{\alpha}\mathcal{R}^{\alpha}$
the Ricci scalar. Therefore, the WL can be written \emph{(up to an
exact differential)} as
\[
\mathcal{L}=\frac{1}{2}\mathcal{R\star}1=-\frac{1}{2}\mathcal{R}_{\alpha\beta}\wedge\star\left(g^{\alpha}\wedge g^{\beta}\right)\text{.}
\]

\end{lemma}

Before we start our geometrization, lets prove the second equality
of the latter equation. Indeed, by Eq.(\ref{Ricci}), $\mathcal{R}_{\alpha}=-i_{\beta}\mathcal{R}_{~\alpha}^{\beta}$,
so that
\[
\mathcal{R}=-i_{\alpha}i_{\beta}\mathcal{R}^{\beta\alpha}=-i_{g_{\beta}\wedge g_{\alpha}}\mathcal{R}^{\beta\alpha}=-i_{g^{\alpha}\wedge g^{\beta}}\mathcal{R}^{\alpha\beta}\text{,}
\]
and therefore
\[
\mathcal{R\star}1=-\mathcal{\star}\left(i_{g^{\alpha}\wedge g^{\beta}}\mathcal{R}^{\alpha\beta}\right)=-\mathcal{R}_{\alpha\beta}\wedge\star\left(g^{\alpha}\wedge g^{\beta}\right)\text{.}
\]
We remark that the above minus sign derives from our definition of
the Ricci tensor.

\bigskip{}

Now, let $(\theta_{~\beta}^{\alpha})$ be the Levi-Civita connection
$1$--forms of $(M,g)$ relative to $\left(g_{\alpha}\right)$. Recalling
the first Cartan structural equation, $dg_{\alpha}=-\theta_{\alpha}^{~\beta}\wedge g_{\beta}$,
we can prove that
\begin{align*}
 & 2dg_{\alpha}\wedge\star dg^{\alpha}-\frac{1}{2}g_{\alpha}\wedge dg^{\alpha}\wedge\star\left(g_{\beta}\wedge dg^{\beta}\right)\\
 & =-\left(\theta_{\alpha\gamma}\wedge g^{\gamma}\right)\wedge\star dg^{\alpha}-\left(\theta_{\alpha\gamma}\wedge g^{\gamma}\right)\wedge\star dg^{\alpha}+\frac{1}{2}g^{\alpha}\wedge\left(\theta_{\alpha\gamma}\wedge g^{\gamma}\right)\wedge\star\left(g_{\beta}\wedge dg^{\beta}\right)\\
 & =-\theta_{\alpha\gamma}\wedge\star dg^{\alpha}\wedge g^{\gamma}+\theta_{\alpha\gamma}\wedge\star dg^{\gamma}\wedge g^{\alpha}-\frac{1}{2}\theta_{\alpha\gamma}\wedge\star\left(g_{\beta}\wedge dg^{\beta}\right)\wedge g^{\alpha}\wedge g^{\gamma}\\
 & =-\theta_{\alpha\gamma}\wedge\star^{2}\left(\star dg^{\alpha}\wedge g^{\gamma}\right)+\theta_{\alpha\gamma}\wedge\star^{2}\left(\star dg^{\gamma}\wedge g^{\alpha}\right)-\frac{1}{2}\theta_{\alpha\gamma}\wedge\star^{2}\left(\star\left(g_{\beta}\wedge dg^{\beta}\right)\wedge g^{\alpha}\wedge g^{\gamma}\right)\\
 & =-\theta_{\alpha\gamma}\wedge\star i^{\gamma}\left(\star^{2}dg^{\alpha}\right)+\theta_{\alpha\gamma}\wedge\star i^{\alpha}\left(\star^{2}dg^{\gamma}\right)-\frac{1}{2}\theta_{\alpha\gamma}\wedge\star i^{\gamma}i^{\alpha}\star^{2}\left(g_{\beta}\wedge dg^{\beta}\right)\\
 & =\theta_{\alpha\gamma}\wedge\star i^{\gamma}dg^{\alpha}-\theta_{\alpha\gamma}\wedge\star i^{\alpha}dg^{\gamma}-\frac{1}{2}\theta_{\alpha\gamma}\wedge\star i^{\gamma}i^{\alpha}\left(g_{\beta}\wedge dg^{\beta}\right)\\
 & =\theta_{\alpha\gamma}\wedge\star\left[i^{\gamma}dg^{\alpha}-i^{\alpha}dg^{\gamma}+\frac{1}{2}i^{\alpha}i^{\gamma}\left(g_{\beta}\wedge dg^{\beta}\right)\right]\text{.}
\end{align*}
On the other hand, by Eq.(\ref{Levi-Civita}) (also, cf. Lemma \ref{Decomp. formula}),
\[
\theta^{\alpha\gamma}=i^{\gamma}dg^{\alpha}-i^{\alpha}dg^{\gamma}+\frac{1}{2}i^{\alpha}i^{\gamma}\left(g_{\beta}\wedge dg^{\beta}\right)\text{,}
\]
so that
\begin{equation}
2dg_{\alpha}\wedge\star dg^{\alpha}-\frac{1}{2}g_{\alpha}\wedge dg^{\alpha}\wedge\star\left(g_{\beta}\wedge dg^{\beta}\right)=\theta_{\alpha\gamma}\wedge\star\theta^{\alpha\gamma}\text{.}\label{B1}
\end{equation}
Also,
\begin{align}
-dg_{\alpha}\wedge\star dg^{\alpha} & =-\left(-\theta_{\alpha\gamma}\wedge g^{\gamma}\right)\wedge\star\left(-\theta_{~\delta}^{\alpha}\wedge g^{\delta}\right)\nonumber \\
 & =-\left(\theta_{\gamma\alpha}\wedge g^{\gamma}\right)\wedge\star\left(-\theta_{~\delta}^{\alpha}\wedge g^{\delta}\right)\nonumber \\
 & =\theta_{\alpha\gamma}\wedge g^{\alpha}\wedge\star\left(\theta_{~\delta}^{\gamma}\wedge g^{\delta}\right)\text{,}\label{B2}
\end{align}
and, using the proof of Lemma (\ref{Codiff.}),
\begin{align}
-\delta g_{\alpha}\wedge\star\delta g^{\alpha} & =-\delta g_{\alpha}\wedge\star\left(-\star d\star g^{\alpha}\right)=\delta g_{\alpha}\wedge\star^{2}\left(d\star g^{\alpha}\right)\nonumber \\
 & =-\delta g_{\alpha}\wedge d\star g^{\alpha}=\star\left(d\star g_{\alpha}\right)\wedge d\star g^{\alpha}\nonumber \\
 & =d\star g^{\alpha}\wedge\star\left(d\star g_{\alpha}\right)\nonumber \\
 & =\theta_{~\gamma}^{\alpha}\wedge\star g^{\gamma}\wedge\star\left(\theta_{\alpha\delta}\wedge\star g^{\delta}\right)\text{.}\label{B3}
\end{align}
So, by Eqs.(\ref{B1}), (\ref{B2}) and (\ref{B3}), the EHL becomes
\begin{align}
2\mathcal{L} & =\left[2dg_{\alpha}\wedge\star dg^{\alpha}-\frac{1}{2}\left(g_{\alpha}\wedge dg^{\alpha}\right)\wedge\star\left(g_{\beta}\wedge dg^{\beta}\right)\right]\nonumber \\
 & -dg_{\alpha}\wedge\star dg^{\alpha}-\delta g_{\alpha}\wedge\star\delta g^{\alpha}\nonumber \\
 & =\theta_{\alpha\gamma}\wedge\star\theta^{\alpha\gamma}+\theta_{\alpha\gamma}\wedge g^{\alpha}\wedge\star\left(\theta_{~\delta}^{\gamma}\wedge g^{\delta}\right)+\theta_{~\gamma}^{\alpha}\wedge\star g^{\gamma}\wedge\star\left(\theta_{\alpha\delta}\wedge\star g^{\delta}\right)\text{.}\label{B4}
\end{align}
Now, the third term can be written as
\begin{align}
 & \theta_{~\gamma}^{\alpha}\wedge\star g^{\gamma}\wedge\star\left(\theta_{\alpha\delta}\wedge\star g^{\delta}\right)\nonumber \\
 & =\star\theta_{~\gamma}^{\alpha}\wedge g^{\gamma}\wedge\star\left(\star\theta_{\alpha\delta}\wedge g^{\delta}\right)\nonumber \\
 & =-\star^{2}\left(\star\theta_{~\gamma}^{\alpha}\wedge g^{\gamma}\right)\wedge\star\left(\star\theta_{\alpha\delta}\wedge g^{\delta}\right)\nonumber \\
 & =-\star\left(i^{\gamma}\theta_{~\gamma}^{\alpha}\wedge i^{\delta}\theta_{\alpha\delta}\right)\text{,}\label{B5}
\end{align}
while the first two as
\begin{align}
 & \theta_{\alpha\gamma}\wedge\star\theta^{\alpha\gamma}+\theta_{\alpha\gamma}\wedge g^{\alpha}\wedge\star\left(\theta_{~\delta}^{\gamma}\wedge g^{\delta}\right)\nonumber \\
 & =\theta_{\alpha\gamma}\wedge\star\theta^{\alpha\gamma}+\theta_{\alpha\gamma}\wedge g^{\alpha}\wedge i^{\delta}\star\theta_{~\delta}^{\gamma}\nonumber \\
 & =\theta_{\alpha\gamma}\wedge\star\theta^{\alpha\gamma}-i^{\delta}\left(\theta_{\alpha\gamma}\wedge g^{\alpha}\right)\wedge\star\theta_{~\delta}^{\gamma}\nonumber \\
 & =\theta_{\alpha\gamma}\wedge\star\theta^{\alpha\gamma}-\theta_{\alpha\gamma}\wedge\star\theta^{\alpha\gamma}-i^{\delta}\theta_{\alpha\gamma}\wedge g^{\alpha}\wedge\star\theta_{~\delta}^{\gamma}\nonumber \\
 & =i^{\delta}\theta_{\alpha\gamma}\wedge\star\theta_{~\delta}^{\gamma}\wedge g^{\alpha}\nonumber \\
 & =-i^{\delta}\theta_{\alpha\gamma}\wedge\star^{2}\left(\star\theta_{~\delta}^{\gamma}\wedge g^{\alpha}\right)\nonumber \\
 & =-i^{\delta}\theta_{\alpha\gamma}\wedge\star i^{\alpha}\theta_{~\delta}^{\gamma}\nonumber \\
 & =\star\left(i^{\delta}\theta_{\alpha\gamma}\wedge i^{\gamma}\theta_{~\delta}^{\alpha}\right)\text{.}\label{B6}
\end{align}
Therefore, by Eqs.(\ref{B4}), (\ref{B5}) and (\ref{B6}),
\begin{align*}
2\mathcal{L} & \mathcal{=}\star\left(i^{\delta}\theta_{\alpha\gamma}\wedge i^{\gamma}\theta_{~\delta}^{\alpha}-i^{\gamma}\theta_{~\gamma}^{\alpha}\wedge i^{\delta}\theta_{\alpha\delta}\right)\\
 & =-i_{g^{\gamma}\wedge g^{\delta}}\left(\theta_{\alpha\gamma}\wedge\theta_{~\delta}^{\alpha}\right)\star1\\
 & =-\left(\theta_{\alpha\gamma}\wedge\theta_{~\delta}^{\alpha}\right)\wedge\star\left(g^{\gamma}\wedge g^{\delta}\right)
\end{align*}
or simply that
\[
\mathcal{L=~}\frac{1}{2}\theta_{\gamma\alpha}\wedge\theta_{~\delta}^{\alpha}\wedge\star\left(g^{\gamma}\wedge g^{\delta}\right)\text{.}
\]

However,
\begin{align*}
 & d\theta_{\gamma\delta}\wedge\star\left(g^{\gamma}\wedge g^{\delta}\right)\\
 & =d\left(\theta_{\gamma\delta}\wedge\star\left(g^{\gamma}\wedge g^{\delta}\right)\right)+\theta_{\gamma\delta}\wedge d\star\left(g^{\gamma}\wedge g^{\delta}\right)\\
 & =d\left(...\right)+\theta_{\gamma\delta}\wedge\left[-\theta_{~\epsilon}^{\gamma}\wedge\star\left(g^{\epsilon}\wedge g^{\delta}\right)-\theta_{~\epsilon}^{\delta}\wedge\star\left(g^{\gamma}\wedge g^{\epsilon}\right)\right]\\
 & =d\left(...\right)-2\theta_{\gamma\alpha}\wedge\theta_{~\delta}^{\alpha}\wedge\star\left(g^{\gamma}\wedge g^{\delta}\right)\text{.}
\end{align*}
Now, the reader must remember the second Cartan structural equation,
so that the above equation yields
\begin{align*}
\mathcal{R}_{\gamma\delta} & =d\theta_{\gamma\delta}\wedge\star\left(g^{\gamma}\wedge g^{\delta}\right)+\theta_{\gamma\alpha}\wedge\theta_{~\delta}^{\alpha}\wedge\star\left(g^{\gamma}\wedge g^{\delta}\right)\\
 & =d\left(...\right)-2\theta_{\gamma\alpha}\wedge\theta_{~\delta}^{\alpha}\wedge\star\left(g^{\gamma}\wedge g^{\delta}\right)+\theta_{\gamma\alpha}\wedge\theta_{~\delta}^{\alpha}\wedge\star\left(g^{\gamma}\wedge g^{\delta}\right)\\
 & =d\left(...\right)-\theta_{\gamma\alpha}\wedge\theta_{~\delta}^{\alpha}\wedge\star\left(g^{\gamma}\wedge g^{\delta}\right)\text{.}
\end{align*}
Therefore, in terms of curvature,
\[
\mathcal{L}=\frac{1}{2}d\left(...\right)-\frac{1}{2}\mathcal{R}_{\alpha\beta}\wedge\star\left(g^{\alpha}\wedge g^{\beta}\right)\text{. }\blacksquare
\]

\end{document}